\documentclass{rQUF2e}

\usepackage{epstopdf}% To incorporate .eps illustrations using PDFLaTeX, etc.
\usepackage{subfigure}% Support for small, `sub' figures and tables

\theoremstyle{plain}

\theoremstyle{definition}

\theoremstyle{remark}

\usepackage{color}
\usepackage{multirow}
\usepackage[normalem]{ulem}
\newcommand{\pd}[2]{\frac{\partial #1}{\partial #2}}
\newcommand{\bo}[1]{\mathbf{#1}}
\usepackage{verbatim}
\usepackage{subfigure}
\usepackage{natbib}
\usepackage{amsfonts}
\usepackage{amsmath}
\usepackage{graphicx}

\begin{document}

%\jvol{00} \jnum{00} \jyear{2014} \jmonth{October}

\title{Option Pricing Accuracy for Estimated Heston Models}

\author{R. Azencott$\dag$, Y. Gadhyan$^{\ast}$${\ddag}$\thanks{$^\ast$Corresponding author.
Email: yutheeka@gmail.com} and R. Glowinski${\ddag}$\\
\affil{$\dag$ Department of Mathematics, University of Houston, Houston, Texas, USA\\
and Ecole Normale Superieure, Cachan, France\\
$\ddag$ Department of Mathematics, University of Houston,
Houston, Texas, USA} %\received{v2.1 released October 2014}
}

\maketitle

\begin{abstract}
We consider assets for which price $X_t$ and squared volatility $Y_t$ are jointly 
driven by Heston joint stochastic differential equations (SDEs). When the parameters
of these SDEs are estimated from $N$ sub-sampled data $(X_{nT}, Y_{nT})$, estimation 
errors do impact the classical option pricing PDEs. We estimate these option pricing 
errors by combining numerical evaluation of estimation errors for Heston SDEs parameters 
with the computation of option price partial derivatives with respect to these SDEs parameters. This is achieved by solving six parabolic PDEs with adequate boundary conditions. To implement this approach, we also develop an estimator $\hat \lambda$ for the market price of volatility risk, and we study the sensitivity of option pricing to estimation errors affecting $\hat \lambda$. We illustrate this approach by fitting Heston SDEs to 252 daily joint observations of the S\&P 500 index and of its approximate volatility VIX, and by numerical applications to European options written on the S\&P 500 index.\end{abstract}


\begin{classcode}Please provide at least one JEL Classification code\end{classcode}

\section{Introduction}
Option based hedging relies on accurate pricing of option contracts, generally computed after modeling the joint dynamics of the underlying asset price $X_t$ and squared volatility $Y_t$. The shortcomings of Black-Scholes models (\cite{black1973pricing}) have been well identified (see \cite{black1976studies},\cite{melino1990pricing}, \cite{stein1989overreactions} ), and have led to studies of a wide range of stochastic volatility models (e.g. \cite{bates1996jumps}, \cite{heston1993closed}, \cite{hull1987pricing}, \cite{JFQ:4491076}, \cite{CambridgeJournals:4490812}, \cite{stein1991stock},\cite{ wiggins1987option}).
For option pricing, one needs to estimate the parameters of the stochastic dynamics driving $X_t$ and $Y_t$. We study here the option pricing errors due to the parameter estimation errors induced by model fitting to actual data, since these errors can indeed be sizeable for small or moderately large market data sets. \\
This paper focuses on computing the impact of parameter estimation errors on European option pricing, when $X_t$ and $Y_t$ are driven by the classical Heston joint stochastic differential equations (SDEs). The Heston model (\cite{heston1993closed}), which has often been applied to concrete market data, does enable the numerical computation of option prices, either by Fourier inversion ( \cite{carr1999option}) or by solving directly the well-known option pricing PDE (\cite{achdou2005computational}, \cite{heston1993closed}).\\
To fit Heston model parameters to data, we use discretized maximum likelihood parameter estimators, as developed and studied in \cite {azencott2009accurate} and \cite{azencottestimation}. These estimators have explicit closed form expressions, combining sub-sampled data $X_{nT}$ and $Y_{nT}$. In practice volatility ``data" are not directly available, and are naturally replaced by well-established estimates, such as ``implied volatility" or ``realized volatility". \\
Volatility estimation has been intensively studied, often in combination with parameter estimation for stochastic volatility models. We refer for instance to \cite{ait2007maximum} and to papers such as \cite{andersenhandbook}, \cite{atiya2009analytic}, \cite{bollerslev2002estimating}, \cite{broto2004estimation}, \cite{gallant1996moments}, \cite{genon1999parameter}, \cite{jacquier2002bayesian}, \cite{kim1998stochastic} and \cite{shephard2005stochastic}. Parameter estimation based on both asset and option prices data has been explored in \cite{avellaneda2003application}, \cite{bakshi1997empirical}, \cite{chernov2000study}, \cite{duffie2000transform}, \cite{fouque2000mean} and  \cite{pan2002jump}
See also our companion studies (see \cite{AzencottRenTimofeyev}, \cite{AzencottBeriRenTimofeyev}, \cite{RenThesis}) which analyze the impact of replacing true volatilities by realized volatilities in large classes of consistent estimators of the Heston parameters. \\
Heston SDEs have often been used to model actual intra-day data, even though they do not model volatility behaviour at very fine time scales. Indeed they generate log-volatility trajectories with Holder exponent close to $ H = 1/2$, but (see \cite{gatheral2014volatility}), for  many assets  and any  $q >0$,  actual estimates of   $E( |log(Y_t)- log(Y_0) |^q$ are of the order of  $t^{H q}$ with   $H << 1/2$. This has led (see \cite{comte1998long} and \cite{gatheral2014volatility} )  to model   log-volatility dynamics by  stationary  Ornstein-Uhlenbeck processes driven by  a fractional Brownian motion  with small Hurst exponent $ H  \leq 0.2$. Our study could be extended to these types of log-volatility dynamics, but at the cost of several technical complications, including a detailed accuracy analysis for model coefficients estimations. So we  have deliberately restricted our paper to the simpler  Heston SDEs models.\\
We consider generic European options on assets for which price and volatility are driven by Heston joint SDEs. The parabolic PDEs verified by option prices involve four Heston model parameters as well as the unknown market price of volatility risk. We define option price sensitivities to these five parameters through partial derivatives of the option price with respect to these parameters. We derive the five PDEs and boundary conditions satisfied by the option price sensitivities, and we outline the efficient numerical schemes we have implemented to compute these sensitivities. We present our discretized maximum likelihood estimators for SDEs parameters, as well as an estimation technique for the market price of volatility risk. We then indicate how to quantify and compute the impact of parameter estimation errors on option pricing.\\
Finally, we illustrate our approach by analyzing market data for options based on the S\&P 500 index, using the VIX index as a proxy for the S\&P 500 volatility. 
\section{Heston Stochastic Volatility Model}\label{Model}
Let $X_t$ be the asset price at time $t \geq 0 $. The squared instantaneous volatility $Y_t$ of the returns process is defined by $Y_t dt = var (dX_t/X_t)$.\\
In the classical Heston model (\cite{heston1993closed}), the pair $\{X_t,Y_t\}$ is a progressively measurable stochastic process defined on a probability space $(\Omega, \mathcal{F}_t, P)$, endowed with an increasing filtration $\mathcal{F}_t$, and is driven by the following system $\mathcal{H}$ of SDEs under the market measure $P$,
\begin{eqnarray}\label{HestonSDE}
dX_t &=& \mu X_t dt + \sqrt{Y_t} X_t d W_t, \label{stock}\\
dY_t &=& \kappa(\theta - Y_t)dt + \gamma \sqrt{Y_t}d B_t. \label{vol} 
\end{eqnarray}
Here $W_t$ and $B_t$ are standard Brownian motions on $\mathbb{R}$, adapted to the filtration $\mathcal{F}_t$, and have constant instantaneous correlation $\rho$, so that $E[dW_t dB_t] = \rho dt$. The parameter vector 
$ \Theta= ( \kappa,\;\theta,\;\gamma,\;\rho )$ is required to verify the well-known constraints 
\begin{equation} \label{pardomain}
|\rho| <1 , \quad \kappa > 0, \quad \theta >0, \quad \gamma > 0 , \quad 2\kappa\theta > \gamma^2, 
\end{equation} 
where the last inequality ensures the almost sure positivity of $Y_t$ as soon as $Y_0 > 0$. There are no constraints on the drift parameter $\mu$, which as is well known does not appear at all in option pricing equations. $Y_t$ is the ``mean reverting" process studied by Feller (\cite{feller1951two}) and originally used in (\cite{cox1985theory}) to model short-term interest rates. In practice $\Theta$ is unknown and has to be estimated from market data, which are usually sub-sampled at $N+1$ successive times $n T$ starting with $n=0$, for some fixed $T >0$ and positive integer $N$. Moreover while the stock price $X_t$ is directly observable in the market, the squared volatility $Y_t$ is not directly available and has to be estimated, generally by ``realized volatility" or by ``implied volatility". From a theoretical point of view, our companion studies (see \cite{AzencottRenTimofeyev}, \cite{AzencottBeriRenTimofeyev}, \cite{RenThesis}) provide an accuracy analysis for the approximation by realized volatilities.
\subsection{Option pricing PDE}
A European call option based on an asset $A$ is a contract signed at time $t=0$, which fixes a strike or exercise price $K$ and a maturity time or exercise date $\tau$. At maturity time, the option holder has the option to buy at price $K$, from the option writer, one share of asset $A$.\\
Call $X_t$ the price of asset $A$ at time $t$. The pay-off of the option at maturity time is then given by $\Psi(X_{\tau})$ where the \textit{pay-off function} $\Psi(x)$ is defined for $ x > 0 $ by
\[
\Psi(x) = (x - K)^+ = \max(x-K,0).
\]
We systematically assume that the price $X_t$ and the squared volatility $Y_t$ of $A$ are driven by the joint Heston SDEs \eqref{stock} \eqref{vol} with coefficients verifying \eqref{pardomain}.
European call option prices based on asset $A$ verify then a well-known parabolic PDE associated to the elliptic second order differential operator $\mathcal{L}$ defined for $x>0, y>0$ by 
\begin{equation}\label{L}
\mathcal{L} = \frac{1}{2}x^{2}y \partial _{x} ^2 + 
\frac{1}{2}\gamma^{2}y \partial _{y}^2 
+ \rho \gamma x y \partial _{x} \partial _{y} 
+ r x \partial _x 
+ [ \kappa(\theta - y)- \lambda  \gamma \sqrt{y} ] \partial _y - r 
\end{equation} 
The coefficients of $\mathcal{L}$ involve the known risk free rate of return $r$, and the four unknown parameters $ \kappa,\theta,\gamma,\rho $ of the Heston SDEs. Note that the drift parameter $\mu$ of these SDEs does not appear in $\mathcal{L}$ .\\
As in \cite{heston1993closed} and \cite{ikonen2004operator}, the classical coefficient 
$\lambda  \gamma \sqrt{y}$ of $\partial_y$ in $\mathcal{L}$ involves the {\em market price of volatility risk} $\lambda \geq 0$, which, as in Heston's classical presentation, is here taken to be  a deterministic constant, which theoretically  remains the same for all European call options based on the fixed asset $A$ (see \cite{bjrk2009arbitrage}). 
\\
The emergence of $\lambda$ in $\mathcal{L}$ and the associated option pricing PDE is due to the fact that the squared volatility $Y_t$ is typically a \textit{non tradable} asset (see \cite{bjrk2009arbitrage} ). Several empirical studies have shown that $\lambda$ cannot be neglected ( \cite{bakshi2003delta}, \cite{bollerslev2011dynamic}, \cite{buraschi2001price}, \cite{carr2009variance}, \cite{eraker2008volatility}, \cite{fouque2000mean}, \cite{lamoureux1993forecasting}). Since our numerical applications below involve only short term options, we have safely assumed, as in \cite{fouque2000mean}, that the unknown market price of risk $\lambda$ is a constant which does not depend on time. Since $\lambda$ is unknown, we do estimate $\lambda$ from market data, as outlined in section \ref{lambdaest} below.\\
As is well known (see \cite{heston1993closed}), the option price $\mathcal{Z}_t $ is of the form 
\begin{equation*} 
\mathcal{Z}_t = g(X_t, Y_t, \tau - t), 
\end{equation*}
where $g(x,y,t)$ is of class 2 in $(x,y)$ and class 1 in $t$, and is the unique solution of the parabolic PDE 
\begin{equation} \label{PDE}
[ \partial _t - \mathcal{L} ] g = 0 \quad \text{for} \; 
0 < x , 0 < y , 0 < t < \tau,
\end{equation}
verifying on the boundary $\partial G$ of $G = R^{+} \times R^{+} \times [ 0, \tau]$ the following boundary conditions 
\begin{eqnarray}\label{f.boundary}
g(x,y,0) = \Psi(x)= (x-K)^+ &\mbox{for} \; 0 < x , 0 < y, \label{f.boundaryinit}\\
g(0,y,t) = 0 &\mbox{ for } \; 0 < y , 0 < t \leq \tau, \label{f.boundarydirich}\\
\left[ \partial _t - rx \partial _ x - \kappa\theta \partial _y + r \right]g (x,0,t) = 0 & \mbox{for} \; 0< x , 0 < t \leq \tau, \label{f.bdryaty}\\
\lim_{x \to \infty} \partial _{x} g(x,y,t) = 1 & \mbox{for} \; 
0 < y , 0 <t \leq \tau, \label{f.boundaryneumx} \\
\lim_{y \to \infty} \partial _{y} g(x,y,t) = 0 & \mbox{for} \; 
0 < x , 0 < t \leq \tau. \label{f.boundaryneumy}
\end{eqnarray}
The first condition asserts that at maturity time, the option price is equal to the option pay-off. The second condition states that when the asset price $x=0$, the option price must also be $0$. The third condition is the formal limit, as $y \to 0$, of the PDE \eqref{PDE}. Indeed note that, as $y \to 0$, the elliptic operator $\mathcal{L}$ does degenerate into the first order differential operator $D = rx \partial _ x + \kappa\theta \partial _y - r$. \\
The fourth and fifth boundary conditions require the option price to be approximately linear in the asset price $x$ for large $x$, and to be roughly independent of the squared volatility $y$ for large $y$.\\
Four of the boundary conditions are standard, namely the initial value \eqref{f.boundaryinit}, the Dirichlet condition \eqref{f.boundarydirich}, and the two Neumann boundary conditions \eqref{f.boundaryneumx} and \eqref{f.boundaryneumy}. But the boundary condition \eqref{f.bdryaty} at $y=0$ is of a more general type (see \cite{glowinski2003handbook}).
\subsection{Option price as a function of the Heston SDE parameters}
The coefficients of the elliptic operator $\mathcal{L}$ are explicitly  determined by the parameter vector $p = [ \kappa, \theta, \gamma, \rho ,\lambda ]$. Due to the specificity of the 3rd boundary condition in \eqref{f.boundaryinit} - \eqref{f.boundaryneumy} and to the degenerescence of the elliptic operator $\mathcal{L}$ on the boundary $y=0$, classical generic results on parabolic PDEs (see \cite{singler2008differentiability}, \cite{yu2003differentiability}) do not directly imply the differentiability of $g(x,y,t) $ with respect to the vector $p$. To prove differentiability of $g$ with respect to $p$, we now naturally use the Heston closed formulas for $g(x,y,t) $.
\\
The option price $g(x,y,t)$ is actually a function of  $x, y, t, p, \tau , K$. For $x, y, t, p, \tau $ fixed, Heston showed that one can write $g(x,y,t)$ as a linear combination of two  inverse Fourier transforms
\begin{equation} \label{gFourier}
g(x,y,t)  = \frac{1}{2 } + 
\int_0^{+ \infty} Re \{\frac{e^{i z \log(K)}}{i \pi z} ( x e^{F_1(z)} - K e^{F_2(z)} ) \} dz  
\end{equation}
where $F_1(z)$ and $F_2(z)$ are explicit complex valued functions of $z$ and  $x, y, t, p, \tau $ (see \cite{heston1993closed}). Heston's formulas for $F_1$ and $F_2$   involve sums of complex values rational fractions in  $z, x, y, t, p, \tau $ and of square roots of similar complex valued rational   fractions.
\\
The option price $g(x,y,t)$ can be computed by numerical implementation of equation \eqref{gFourier} because the two  integrals in $z >0$ do converge fast enough as 
$z \to  + \infty$. But in this paper, we are specifically focused on quantifying option price sensitivities to estimation errors  on  the parameter vector $p$, a goal which  requires computing the gradient of $g(x,y,t)$with respect to $p $, denoted by
\begin{equation} \label{partial_p}
\partial_{p} g = [ \partial_{\kappa} g , \partial_{\theta} g , \partial_{\gamma} g , \partial_{\rho} g , \partial_{\lambda} g ]
\end{equation}
Formally, equation \eqref{gFourier} yields the formula 
\begin{equation} \label{partial.gFourier}
\partial_{p}g(x,y,t) = 
\int_0^{+ \infty} Re \{\frac{e^{i z \log(K)}}{i \pi z} \partial_{p} ( x e^{F_1(z)} - K e^{F_2(z)} ) \} dz  
\end{equation}
The  gradients  $ \partial_{p}  e^{F_1(z)}$ and  $ \partial_{p}  e^{F_2(z)}$ can be computed explicitly and  this yields 10 quite cumbersome formulas for the 5  partial derivatives of $e^{F_1(z)}$ and  those of $e^{F_2(z)}$. Each such formula is a function of   $z, x, y, t, p, \tau$ involving compositions of exponentials, square roots, products, and sums of complex valued rational fractions.  
\\
We have checked the  absolute convergence  of the 10  integrals in $z$ involved in   equation \eqref{partial.gFourier}, in particular by verifying that the modulus of their 10 integrands is for large $z >0 $ necessarily inferior to $z^{\alpha }e^{- \beta z}$ for some positive  $\alpha$ and  $\beta$ determined by $x,y,t,p, \tau$. These extensive but tedious verifications have led us in fine to the validation of formula \eqref{partial.gFourier}. 
\\
On the practical level, the full reliability of these formulas is slightly impaired due to the multiple ambiguities generated by simultaneously selecting the determinations of many square roots of complex rational functions (and of their derivatives). Making sure that one chooses the proper combinations of such square roots determinations  in all parameter configurations is not a standard computing task, and this point  led us to avoid computing $\partial_{p}g$ by numerical implementation of  equation \eqref{partial.gFourier}. 
\\
So for concrete numerical computations of the option price gradient $\partial_{p}g$ with respect to the parameter vector $p$, we have definitely preferred to derive and solve  the 5 parabolic PDEs \eqref{sensitivity_eqns} verified by the 5 coordinates of $\partial_{p}g$. This generic approach has the merit of extending easily to parameterized pairs of   driving SDEs much more general than the Heston SDEs. Moreover the 5 PDE solvers used here are essentially similar and converge reasonably fast as seen below.
\section{Option price sensitivity to parameters estimation errors}
%%%%%%%%%%%
%
\subsection{Option price sensitivity : definition } \label{def.sensitivity}
In practical option pricing, the parameter vector $p$ has to be estimated from price data, approximate volatility data, and/or observed option trading prices. The unavoidable estimation errors on $p$ necessarily induce option pricing errors. \\
We define the five {\em sensitivities} of the option pricing function $ f(x,y,t) = g(x,y,\tau -t)$ with respect to small errors affecting the parameter vector $p$ by the formulas 
\begin{equation} \label{defSen}
Sen_{\kappa} = | \partial_{\kappa}g | \quad Sen_{\theta} = | \partial_{\theta}g | \quad Sen_{\gamma} = | \partial_{\gamma}g | \quad Sen_{\rho} = | \partial_{\rho}g| \quad Sen_{\lambda} = | \partial_{\lambda}g | \quad 
\end{equation}
where all partial derivatives are computed at the point $(x,y,\tau - t)$. These option price sensitivities are hence functions of $(x,y,t, p, \tau , K )$. The numerical computation of option price sensitivities has been explored in \cite{broadie2004exact, chan2010first} and we propose here an alternate approach. 
%%%%%%%%
\subsection{Option Price Sensitivity PDEs }\label{sensitivity}
%%%%%%%%
%%
Differentiating equation \eqref{PDE} with respect to each one of the coordinates of the parameter vector $p$, one obtains the five independent PDEs verified by the 5 coordinates of the gradient $\partial_{p} g (x,y,t)$, in the open domain $0 < x ,\; 0 < y ,\; 0 < t < \tau$. 
\begin{eqnarray} \label{sensitivity_eqns}
( \pd{}{t} - \mathcal{L} ) \, \partial_{\kappa} g &= & (\theta - y)\pd{g}{y} \label{sens_kappa} \\
( \pd{}{t} - \mathcal{L} ) \, \partial_{\theta} g &= & \kappa\pd{g}{y} \label{sens_theta}\\
( \pd{}{t} - \mathcal{L} ) \, \partial_{\gamma} g & = & \gamma y \frac{\partial^{2}g}{\partial y^2} + \rho x y \frac{\partial^{2}g}{\partial x\partial y} -\lambda \sqrt{y} \pd{g}{y} \label{sens_gamma}\\
( \pd{}{t} - \mathcal{L} ) \, \partial_{\rho} g &= & \gamma x y \frac{\partial^{2}g}{\partial x \partial y} \label{sens_rho} \\
( \pd{}{t} - \mathcal{L} ) \, \partial_{\lambda} g & = & -\gamma\sqrt{y}\pd{g}{y} \label{sens_lambda}
\end{eqnarray} 
Each one of these five PDEs is associated to five boundary conditions. The first four of these conditions are given in compact form by the following vector equations, where $\mathbf{[ 0 ]}$ is the vector of $R^5$ with all coordinates equal to 0, 
\begin{eqnarray}\label{partial.f.boundary}
\partial_{p} g & = & 0 \mbox{ for } \; 0 <x ,\; 0 < y,\; t = 0 \\
\partial_{p} g & = & 0 \mbox{ for } \; x= 0 ,\; 0 < y ,\; 0 < t \leq \tau \\
\lim_{x \to \infty} \partial _{x} [ \partial_{p} g ] & = & \mathbf{[ 0 ]}\; \mbox{for} \; 0 < y , \; 0 <t \leq \tau \\
\lim_{y \to \infty} \partial _{y} [ \partial_{p} g ] & = & 0\; \mbox{for} \; 0<x ,\; 0 <t \leq \tau 
\end{eqnarray}
The fifth boundary conditions are obtained by differentiating equation \eqref{f.bdryaty} with respect to each one of the parameters. This yields the following boundary conditions, valid for $ 0 < x , \; y = 0 , \; 0 <t \leq \tau $, 
\begin{eqnarray}
\mathcal{D} \, \partial_{\kappa} g = \theta \, \partial_{y} g,\quad
\mathcal{D} \, \partial_{\theta} g = \kappa \, \partial_{y} g,\quad
\mathcal{D} \, \partial_{\gamma} g = 0,\quad
\mathcal{D} \, \partial_{\rho} g = 0, \quad
\mathcal{D} \, \partial_{\lambda} g = 0.
\end{eqnarray}
where $\mathcal{D}$ is the 1st order differential operator given by 
\begin{equation}
\mathcal{D} = \pd{}{t} - rx\pd{}{x} - \kappa \theta \pd{}{y} + r .
\end{equation}
We now outline the numerical scheme we have implemented to solve the option price PDE and the five preceding non homogeneous PDEs.
%%%%%%%%%%%
\section{Numerical computation of option price sensitivities to parametric errors}\label{numerical_imp}
%%%%%%%
In concrete numerical option pricing, the asset price $x > 0$ and the squared volatility $y > 0$ have obvious explicit realistic upper bounds $x_B,y_B$, so it is standard computing practice to replace the unbounded domain $G = R^{+} \times R^{+} \times [0,\tau] $ by the bounded domain $ \; U = [0,x_B ]\times [0,y_B]\times [0,\tau] \; $ where $x_B,y_B$ are fixed large positive numbers. \\
Then $g(x,y,t) $ will be computed as the unique function verifying the parabolic PDE \eqref{PDE} for $(x,y,t)$ in the interior $U^o$ of $U$ and the following five conditions on the boundary $\partial U$ of $U$
\begin{eqnarray}
g(x,y,0) = \Psi(x)= (x-K)^+ \mbox{ for } \; 0 < x < x_B ,\; 0 < y < y_B \label{init} \\
g(0,y,t) = 0 \mbox{ for } \; 0 < y < y_B ,\; 0 < t \leq \tau \label{bdryx0} \\
\left[ \partial _t - rx \partial _ x - \kappa\theta \partial _y + r \right]g (x,0,t) = 0 \; \mbox{for} \; 0< x < x_B ,\; 0 < t \leq \tau \label{bdry} \\
\lim_{x \to x_B } \partial_{x} g(x,y,t) = 1 \; \mbox{for} \; 
0 < y <y_B ,\; 0 <t \leq \tau \label{bdryxinfty} \\
\lim_{y \to y_B } \partial_{y} g(x,y,t) = 0 \; \mbox{for} \; 
0 < x <x_B , \; 0 <t \leq \tau \label{bdryyinfty} 
\end{eqnarray}
To solve the PDE \eqref{PDE}, we discretize the elliptic operator $\mathcal{L}$ in \eqref{L}, by standard numerical schemes well known to be stable under discretization refining. Numerical schemes for option pricing are discussed in multiple papers such as \cite{achdou2005computational}, \cite{ikonen2004operator}. For European options under the Heston model, the papers \cite{doi:10.1080/00207160.2013.777710} and \cite{RePEc:wsi:ijtafx:v:16:y:2013:i:03:p:1350015-1-1350015-35} both present explicit specific numerical schemes.\\
We apply a uniform space-time finite difference grid on the domain $U$ using the second order space discretization outlined in \cite{ikonen2004operator} and the backward differentiation formula for time discretization given in \cite{oosterlee2003multigrid} .\\
Let the number of grid steps be $m,n$ and $s$ in the $x,y$ and $t$ directions respectively. Grid step sizes in each direction are denoted 
\[
\Delta x = x_B / m ; \quad \Delta y = y_B / n ; \quad \Delta t = \tau / s .
\]
At grid points, the values of $g$ are indexed as follows
\[
g_{ij}^{k} = g(x_i,y_j,t_k) = g(i\Delta x,j\Delta y,k\Delta t) \quad \text{for} \quad i = 0,\dots, m ; j=0, \dots, n ; k=0, \dots, s.
\]
\subsection{Space discretization}
All space partial derivatives in the parabolic PDE \eqref{PDE} have variable coefficients, so that in some parts of the domain the first order derivative terms may dominate the second order terms. To discretize these spatial derivatives we apply a space discretization scheme used in \cite{ikonen2004operator} for American options. After discretization, $\mathcal{L}$ becomes a matrix $\bo{A}$ well studied in \cite{ikonen2004operator}. \\
Recall that M-matrices which are strictly diagonally dominant with positive diagonal elements and non-positive off diagonal elements have good stability properties (see \cite{varga-matrix}, \cite{windisch1989m}). In general $\bo{A}$ is not an M-matrix but as remarked in \cite{ikonen2004operator}, when the time discretization of \eqref{PDE} has sufficiently small time steps, $\bo{A}$ becomes diagonally dominant. 
\\
We apply the seven point spatial discretization scheme of \cite{ikonen2004operator} to solve for the option price. We use a second order accurate finite difference scheme for the space derivatives, namely the classical central difference scheme for first order derivatives and the usual three point scheme for second order derivatives.\\ 
The corresponding finite difference operators are then 
\begin{equation}\label{finite_first}
\delta_x g_{i,j}^k = \frac{g_{i+1,j}^k - g_{i-1,j}^k}{2\Delta x}, \quad \delta_y g_{i,j}^k = \frac{g_{i,j+1}^k - g_{i,j-1}^k}{2\Delta y},
\end{equation}
\begin{equation}\label{finite_second}
\delta^2_x g_{i,j}^k = \frac{g_{i+1,j}^k - 2g_{i,j}^k +g_{i-1,j}^k}{\Delta x^2} , \quad \delta^2_y g_{i,j}^k = \frac{g_{i,j+1}^k - 2g_{i,j}^k +g_{i,j-1}^k}{\Delta y^2}.
\end{equation}
On the boundary $j=0$, the derivative in the $y$ direction in \eqref{bdry} is evaluated by the following upwind discretization scheme (see \cite{glowinski2008numerical}),
\[
\delta_y g_{i,0}^{k} = \frac{-3g_{i,0}^{k}+4g_{i,1}^{k}-g_{i,2}^{k}}{2\Delta y}.
\]
As proposed in \cite{ikonen2004operator}, the mixed derivatives are discretized using a seven point stencil $ \delta_{xy}$, where $ 2 \Delta x \; \Delta y \; \delta_{xy} \; g_{i,j}^k $ is given by 
\begin{equation} \label{finite_mixed}
2g_{i,j}^k + g_{i+1,j+1}^k + g_{i-1,j-1}^k - g_{i+1,j}^k - g_{i-1,j}^k - g_{i,j+1}^k - g_{i,j-1}^k .
\end{equation}
%%%
We handle the Neumann boundary conditions \eqref{bdryxinfty} and \eqref{bdryyinfty} in the same way as in \cite{ikonen2004operator}. 
This space discretization leads to a semi-discrete equation,
\[
\frac{d\bo{g}}{dt} + \bo{A}\bo{g} = \bo{B},
\]
where $\bo{A}$ is an $mn \times m n$ matrix and $\bo{B}$ is a column vector of length $m n$. The vector $\bo{g}$ of length $m n$ gathers all the option price values at grid points. The vector $\bo{B}$ gathers terms due to the Neumann boundary condition in the $x$ direction and does not depend on $t$. 
%%%
\subsection{Time discretization}
%%%
For time discretization, as in \cite{oosterlee2003multigrid}, we use the BDF2 scheme (Backward Difference Formula), which is an implicit scheme with second order accuracy \cite{oosterlee2003multigrid}. At time $k\Delta t$ the BDF2 scheme reads,
\[
\dfrac{3\bo{g}^{k+1}-4\bo{g}^{k}+\bo{g}^{k-1}}{2\Delta t} + \bo{A}\bo{g}^{k+1} = \bo{B},
\]
for $k=1,2,\dots,l-1$. 
The stability properties of this scheme are studied in \cite{ikonen2004operator} and \cite{oosterlee2003multigrid}. At each iterate of the BDF2 scheme we require the value of the last two iterates. To obtain the 1st iterate, we use an Implicit Euler scheme \cite{glowinski2008numerical} : given the initial value $\bo{g}^{0}$, we compute $\bo{g}^{1}$ by 
\[
\frac{\bo{g}^{1} - \bo{g}^{0}}{\Delta t} + \bo{A}\bo{g}^{1} = \bo{B} .
\]
At moderate grid sizes, we have numerically verified that this choice of space-time discretization of our initial-boundary value problem gives us stable solutions for the option price. At each time step, we use an LU decomposition to solve the following system of linear equations,
\begin{equation}\label{discrete}
(\bo{I} + \frac{2}{3}\Delta t \bo{A})\bo{g}^{k+1} = \frac{4}{3}\bo{g}^{k} - \frac{1}{3}\bo{g}^{k-1}+\Delta t \bo{B}, 
\end{equation}
where $\bo{I}$ is the $mn \times mn$ identity matrix.\\
Once the function $g$ has been computed on our discrete grid, we solve the sensitivity equations \eqref{sens_kappa}-\eqref{sens_lambda} by a numerical scheme quite similar to the scheme just described. 
To evaluate the right-hand side of equations \eqref{sensitivity_eqns}, we use a central difference scheme to discretize the first spatial derivatives of $g$, a 3-point stencil similar to \eqref{finite_second} for the second spatial derivatives of $g$, and a 7 point stencil similar to \eqref{finite_mixed} for the mixed second derivative of $g$. The space-time discretization of the sensitivity equations is the same as above, and we can then solve these discretized equations on the same grid used to compute the option price.\\
We have verified empirically, by successive grid refinements, that this numerical scheme generates converging approximations for the solutions of the sensitivity equations. 
\section{ Estimation of Parameters for Heston joint SDEs }\label{estim_errors}
Asset prices $X_t$ are directly observed, but squared volatilities $Y_t$ are not directly observable and have to be estimated indirectly. The most common estimators of $Y_t $ are the ``implied" squared volatility derived by analysis of option prices and the ``realized" squared volatility   (see \cite{andersen2003modeling, barndorff2002econometric, dacunha1986estimation, garman1980estimation, genon1994estimation} for estimation of realized volatility).  For our study of S\&P500 daily data below, a standard approximation of $\sqrt{Y_t}$ is provided by a fixed multiple of the $VIX$ index. 
\\
In concrete contexts, the available price data are $N$ observations $U_n = X_{nT} , \; n=1,2,\ldots,N, \;$, where $T$ is a fixed user selected sub-sampling time, with standard values such as $T=1/252$ for daily data. The subsampled squared volatilities $ V_n= Y_{nT} $ are not observable and are  estimated by $\hat{V}_n$, usually computed by squaring either  implied volatilities or  realized volatilities.
Call $ \Theta = ( \kappa,\theta,\gamma^2,\rho )$ the unknown parameter vector of the 
Heston  SDEs  driving $ X_t, Y_t$. In companion preprints (see \cite{AzencottRenTimofeyev}, \cite{AzencottBeriRenTimofeyev}, \cite{RenThesis}) we have studied how the replacement of $V_n$ by squared realized volatilities $\hat{V}_n$ impacts  estimation consistency for the  parameter vector $\Theta$ of the Heston SDEs. These results identify   large classes of asymptotically consistent estimators  $F_N = F_N(V_1, \ldots , V_N)$  of  $\Theta$, such that the associated \textit{observable} estimators $\hat{F}_N = F_N ( \hat{V}_1, \ldots , \hat{V}_N)$ of  $\Theta$ remain asymptotically consistent, provided the window length  $r$ and the subsampling step $T$  are forced to depend on $N$ at specific but explicit polynomial rates. We conjecture that under adequate hypotheses,  asymptotic stability results also hold for the replacement of $V_n$ by squared implied volatilities.
\\
This  type of   asymptotic stability  suggests to estimate $\Theta$ as follows. Start with $N$ observable data ($U_n= X_{nT}, \hat{V}_n= \hat{Y}_{nT})$ subsampled from the asset price $X_t$ and from   estimates $\hat{Y}_t$ of the unknown square volatility $Y_t$ of $X_t$. Concretely $\hat{Y}_t$ is computed either by squared realized volatilities  or by squared implied volatilities.  In \cite{azencottestimation} we have introduced and studied at length explicit \textit{ discretized maximum likelihood estimators} $  F_N(V_1, \ldots , V_N)$  of  $\Theta$.  Asymptotically in N, these estimators are consistent and nearly most efficient, but of course they are not directly observable since they involve the true squared volatilities $V_n$. However in the explicit formulas $F_N$ recalled below, we then replace each $V_n$ by its estimate $\hat{V}_n $, which provides us with \textit{observable} estimators of $\Theta$, of the form 
$$
\hat{\Theta_N} = F_N ( \hat{V}_1, \ldots , \hat{V}_N )
$$
Let us describe how the ``non observable" estimator $ F_N$ is computed from the true volatilities $V_N$  in \cite{azencottestimation}, where it is derived by  likelihood maximization after formal Euler discretization of the Heston joint SDEs. This approach leads to define the five sufficient statistics $a,b,c,d,f$ 
\begin{equation}\label{coeff1}
\begin{split}
a =& \frac{1}{N}\sum_{n=0}^{N-1} \frac{(V_{n+1} - V_n)^2}{V_n} \; , 
\quad b = - \frac{2}{N}\sum_{n=0}^{N-1} \frac{V_{n+1} - V_n}{V_n} \; , \quad c = \frac{2}{N}(V_N -V_0) \\
d =& \frac{2}{N}\sum_{n=0}^{N-1} \frac{1}{V_n} \; ,\quad f = \frac{2}{N}\sum_{n=0}^{N-1} V_n.
\end{split}
\end{equation}
These statistics almost surely verify 
\begin{equation*} 
a>0 \; , \; d > 0 \; , \; f>0 \; ,\quad df - 4 >0 \; , \quad 2 a f - c^2 > 0 \; 
\quad d+f -4 > 0.
\end{equation*} 
Our nearly maximum likelihood estimators of $ \kappa, \theta, \gamma^2 $ are then explicitly given by 
\begin{equation}\label{hatestim}
\hat \kappa = -\frac{2b+cd}{T(df-4)} \; ,\quad \quad \hat \theta = \frac{bf+2c}{2b+cd} \; , \quad \quad \hat \gamma^2 = \frac{a}{T} - \frac{b^2 f+4bc+c^2 d }{2T(df-4)} .
\end{equation} 
When $N\to \infty$ and with $T$ fixed and small enough, we have shown in \cite{azencottestimation} that, with probability tending to 1, these three estimators verify all the required natural constraints \eqref{pardomain} and converge in probability to the true parameters. \\ 
Recall that the Heston SDEs \eqref{stock} - \eqref{vol} involve two Brownian motions $W_t$ and $B_t$. Once $ \kappa, \theta, \gamma$ are estimated, the discretization of the SDEs \eqref{stock} and \eqref{vol} with time step $T$ provides natural estimates $\hat{\mu}$ for $\mu$ and $DZ_n$, $DB_n$ for the Brownian increments $ W_{(n+1)T} - W_{nT}$ and $ B_{(n+1)T} - B_{nT}$. The natural estimator $\hat{\rho}$ of $\rho$ is then the empirical correlation of $DZ_n$ and $DB_n$, which is an explicit rational fraction  involving only $N,T$  and $V_1, V_2, \ldots, V_N$ .
\\
At this point the ``non observable" estimator $F_N = ( \hat{\kappa}, \hat{\theta}, \hat{\gamma}^2, \hat{\rho} ) $ is fully and explicitly specified in terms of $V_1, \ldots, V_N$.
In the preceding formulas giving $a,b, c, d, f$ and $F_N$, we now replace all the $V_n$ by the observable $\hat{V_n}$, to obtain an explicit \textit{observable} estimator $\hat{\Theta}_N$, which is  asymptotically consistent as $N \to \infty$ provided $T$ is fixed and small enough. For brevity below we will often omit the subscript $N$ in $\hat{\Theta}_N$ since the number $N$ of observations is fixed  in our data studies below.
\\
Call $\Delta \Theta$ the vector $( \hat{\Theta} - \Theta ) $ of parameter estimation errors. These theoretical errors are not necessarily in $L_2$, because $1/Y_t$ does not always have a finite second moment but any slight truncation of our estimators eliminates this difficulty in numerical applications to concrete data. The covariance matrix $\Sigma $ of $\Delta \Theta$ provides then the $L_2$-sizes of estimation errors and their correlations. \\
We have validated in ( \cite{azencottestimation}, \cite{gadhyan2010option} ) that for $N$ moderately large, one can generate reasonable empirical estimates $\hat{\Sigma}$ of the error covariance matrix $\Sigma$ as follows. Use the $N$ observations $X_{nT}, Y_{nT}$ to compute the associated value $\Theta_0$ of our vector of estimators $\hat{\Theta}$. Then simulate a moderately large number $q$ of random diffusion trajectories $\omega_j$ of duration $NT$ driven by joint Heston SDEs parameterized by $ \Theta_0 $. This is achieved by a standard Euler time discretization of the Heston SDEs, with very small discretization time step $\delta << T$. The explicit estimation formulas outlined above then provide one value $\Theta_j$ of the random vector $ \hat{\Theta} $ for each simulated trajectory $\omega_j$. The empirical covariance matrix $\hat{\Sigma}$ of the $\Theta_j - \Theta_0$ is then a natural estimator of the covariance matrix $\Sigma$ . \\
%%%%%%
\section{Impact of parametric estimation errors on option pricing }
%%%%%%%%%%% 
To implement option pricing, we start from $N$ joint observations $(X_{nT}, Y_{nT})$ of the underlying asset price and squared volatility, where $T$ is a fixed known (or user selected) sub-sampling time step. As just described, we use these data to compute an estimator $\hat{\Theta}$ for the vector $\Theta$ of Heston model coefficients.\\
Consider a European option based on this asset, with given strike price $K$ and maturity date $\tau$. In the option pricing PDE \eqref{PDE}, the unknown Heston model coefficients are then replaced by the estimators just computed. The option price $f(x,y,t) = g(x,y, \tau -t)$ computed by solving \eqref{PDE} is then affected by a (random) error $\Delta f(x,y,t) $. Our objective is to compute numerical bounds for this option pricing error. The option price 
$f(x,y,t) = g(x,y, \tau -t)$ depends also on the underlying parameter vector $\Theta$ and $\lambda$. For the moment we consider the impact of errors in the parameter vector $\Theta$ only. The impact of estimation errors affecting $\lambda$ will be studied separately below in section \ref{lambdaest}.
To simplify notations we often omit below the variables $(x,y,t, \Theta, \lambda)$ for $f$ and the associated variables $(x,y, \tau -t, \Theta,\lambda)$ for $g$.\\
With this shorthand convention, the option pricing error $\Delta f$ is equal to the error 
$ \Delta g$ affecting $g(x,y, \tau-t)$. For $N T$ large enough and $T$ small enough, the vector of estimation errors $\Delta \Theta = \hat{\Theta} - \Theta $ becomes arbitrarily small (see \cite{azencottestimation}), so that one can legitimately apply a first order Taylor expansion to obtain the approximation 
\begin{equation} \label{deltaf}
\Delta f = \Delta g \; \simeq \; \partial_{\Theta} g . \; \Delta \Theta.
\end{equation}
For each fixed quadruplet $(x,y, t, \Theta)$, parameter estimation errors have a \textit{perturbation impact} on option pricing, and we quantify this impact by the $L_2$-norm $ \varepsilon $ of $ \Delta f $. Due to equation \eqref{deltaf}, we have the approximation 
\begin{equation}\label{impactformula}
\varepsilon^2 \; \simeq \; \partial_{\Theta} g^* . \; \Sigma . \; \partial_{\Theta}g, \; 
\end{equation} 
where the gradient $ \partial_{\Theta}g $ should be evaluated at $(x,y,\tau - t, \Theta)$, and $*$ denotes matrix transpose.
%and can be approximated by its value at $(x,y,\tau - t, \hat{\Theta})$.
Since $\Sigma $ is positive definite, we have $|\Sigma_{i,j}| < \sqrt{\Sigma_{i,i} \Sigma_{j,j}}$ for all $i,\;j \in \{1,2,3,4\}$, which implies, for all $v \in R^4$, the inequality
\begin{equation} \label{v*etav}
0 \leq v^*. \, \Sigma . \, v \leq (\sum_{i=1 \ldots 4} |v_i| \Sigma_{i,i}^{1/2})^2.
\end{equation}
The squared $L_2$-norms $ s_{\kappa}^2 , s_{\theta}^2 , s_{\gamma}^2 , s_{\rho}^2 $ of estimation errors on $ \kappa,\theta, \gamma, \rho $ are the diagonal terms $\Sigma_{i,i}$ of $\Sigma $, so that equations \eqref{v*etav} and \eqref{impactformula} yield the upper bound
\begin{equation} \label{eps}
\varepsilon \; \leq \; s_{\kappa} \, | \partial_{\kappa}g |
+ s_{\theta} \, | \partial_{\theta}g |
+ s_{\gamma} \, | \partial_{\gamma}g|
+ s_{\rho} \, | \partial_{\rho}g|. 
\end{equation}
Introducing the option price sensitivities $Sen_{\kappa} , Sen_{\theta} , Sen_{\gamma} , Sen_{\rho}$ as defined in section \ref{def.sensitivity}, the last inequality becomes 
\begin{equation} \label{eps2}
\varepsilon \; \leq \; s_{\kappa} \, Sen_{\kappa} + s_{\theta} \, Sen_{\theta}
+ s_{\gamma} \, Sen_{\gamma} + s_{\rho} \, Sen_{\rho} 
= \varepsilon_{\kappa} + \varepsilon_{\theta} + \varepsilon_{\gamma} + \varepsilon_{\rho},
\end{equation}
where the individual impacts on option pricing of the estimation errors respectively affecting $\kappa, \theta, \gamma, \rho $ are defined by 
\begin{equation} \label{singleimpact}
\varepsilon_{\kappa} = s_{\kappa} \, Sen_{\kappa} \; , \quad \varepsilon_{\theta} = s_{\theta} \, Sen_{\theta}, \quad \varepsilon_{\gamma} = s_{\gamma} \, Sen_{\gamma} , \quad \varepsilon_{\rho} = s_{\rho} \, Sen_{\rho}.
\end{equation}
%% 
%%%%%%%%%
%\section{Numerical Study of Option pricing accuracy}\label{option_numerics}
%%%%%%% 
%For the stock market example studied below, we fix a subsampling time $T$ and we consider that the $N$ observable data ($U_n= X_{nT}, \hat{V}_n= %\hat{Y}_{nT})$ are subsampled from the asset price $X_t$ and from estimates $\hat{Y}_t$ of the unknown squared volatilities $Y_t$. 
%%%%%
\subsection{Joint SDEs model fitting for S\&P 500 and VIX}
%%%%%%
For the stock market example studied below, we fix a subsampling time $T$ and we consider that the $N$ observable data ($U_n= X_{nT}, \hat{V}_n= \hat{Y}_{nT})$ are subsampled from the asset price $X_t$ and from estimates $\hat{Y}_t$ of the unknown squared volatilities $Y_t$. 
To illustrate how we quantify the impact of parameter estimation errors on option pricing, we study the case of two options written on the index S\&P 500.  
We used a dataset $ (SPX)_n, (VIX)_n$ of $N=252$ joint daily observations recorded in 2006 for the S\&P 500 index and its approximate annualized volatility VIX. Recall that VIX, as maintained by CBOE, estimates SPX volatility through the implied volatility of options with a 30 day maturity, and is annualized on the standard basis of 252 trading days per year (see \cite{vixwhite}). We consider the historical time series of $ (SPX)_n$ and its volatility proxy as being viewed under the market measure $P$.  After fixing a conventional time interval $T=1/252$ between two successive daily observations, the daily SPX observations are denoted by $(SPX)_n = X_{nT}$ where $(X_t, Y_t)$ is an underlying Heston process driven by standard Heston SDEs with unknown parameter vector $\Theta$. 
\\
Given N joint daily data  $  (SPX)_n , (VIX)_n$, we thus set  $U_n = X_{nT} =  (SPX)_n $, and we estimate the unobservable squared volatilities $V_n= Y_{nT}$ by $\hat{V}_n = b \times (VIX)_n ^2$  where the  fixed coefficient $b$ simply takes  account of the   fixed annualization coefficient involved in VIX.
 As explained in section \ref{estim_errors}, to estimate $\Theta$,  we then first write the explicit formulas \eqref{hatestim} giving  our consistent \textit{maximum likelihood estimators}, expressed in terms of the $V_n$ and the $U_n$; in these formulas, we  replace each $V_n$ by the observable estimates $\hat{V}_n $, and this   provides us with \textit{observable} asymptotically consistent estimators $\hat{\Theta}_N$ of $\Theta$. 
\\
From the $N= 252$ data  $ (SPX)_n, (VIX)_n$ recorded in 2006, we generate the time series $(SPX)_n, \hat{V}_n = b \times (VIX)_n ^2$ to which we apply our parameter estimators, as defined in equations \eqref{coeff1} and \eqref{hatestim}. This yields the vector $\hat \Theta $ of estimated parameter  values 
\begin{equation}
\hat\kappa = 16.6 \;;\; \hat\theta = 0.017 \;;\; \hat\gamma = 0.28 \;;\; \hat\rho = -0.54.
\end{equation}
The estimated drift $\mu$ is not listed since it has no impact on option pricing. \\ 
As outlined in section \ref{estim_errors}, we then fix $\Theta$ at the estimated values just obtained to simulate 5000 trajectories of the just fitted Heston SDEs, and this enables the computation of empirical evaluations for the root mean squared errors of estimation $s_\kappa , s_\theta , s_\gamma , s_\rho$. This easily yields the values 
\begin{equation}
\hat s_\kappa \sim 5.7 \;;\; \hat s_\theta \sim 0.002 \;;\; \hat s_\gamma \sim 0.01 \;;\; \hat s_\rho \sim 0.06.
\end{equation}
The number $N= 252 $ of joint daily observations is realistic but rather small, so that the relative errors of estimation on SDEs parameters are naturally still fairly high.\\
For numerical evaluations below, we consider that with reasonably high probability, the parameters $\kappa,\; \theta,\; \gamma,\; \rho$ belong to four intervals centered around 
$\hat \kappa,\; \hat \theta,\; \hat \gamma,\; \hat \rho $, and with half-widths $\hat s_{\kappa},\; \hat s_\theta, \; \hat s_\gamma, \; \hat s_\rho$. The product $J \subset R^4$ of these four intervals is then a high probability ``localization box" for the true vector $\Theta$ of unknown parameters.
%%%%%%%%%%%
\section{An estimator for the market price of volatility risk} \label{lambdaest}
%
%To compute the option price, one first has to estimate the unknown market price $\lambda$ of volatility risk which depends on investors preferences, %liquidity concerns, risk aversion, etc. For various approaches to estimate $\lambda$, see for instance \cite{bjrk2009arbitrage}, \cite{fouque2000mean}, %\cite{heston1993closed}, \cite{lamoureux1993forecasting}.\\
%In the spirit of \cite{fouque2000mean}, and since in our numerical examples we consider only short term options maturing within the same short period of %time, we have safely assumed that $\lambda$ is an unknown constant, and we  use a  small pool of short term options to estimate  $\lambda$  as follows. 
% 
\subsection{Estimation of the  market price  of volatility risk }\label{hatlambda}
To compute the option price, one first has to estimate the unknown market price $\lambda$ of volatility risk which depends on investors preferences, liquidity concerns, risk aversion, etc. For various approaches to estimate $\lambda$, see for instance \cite{bjrk2009arbitrage}, \cite{fouque2000mean}, \cite{heston1993closed}, \cite{lamoureux1993forecasting}.\\
In the spirit of \cite{fouque2000mean}, and since in our numerical examples we consider only short term options maturing within the same short period of time, we have safely assumed that $\lambda$ is an unknown constant, and we  use a  small pool of short term options to estimate  $\lambda$  as follows. \\
Consider $q$ benchmark options $\Omega_j$, $ j = 1 \ldots q $, with strikes $K_j$ and maturity $\tau_j$, written on the same underlying asset. Consider the underlying asset denoted by $X_t$ and its volatility $Y_t$ are driven by SDEs \eqref{stock}-\eqref{vol}. Call $T$ the user selected time step between observations. Then for each day $k$, let $X_{kT}$ be the closing asset price and $\Omega_j(kT)$ be the average of closing bid and ask for $\Omega_j$. We will estimate the unknown constant $\lambda$ by minimizing a distance between the predicted option price and the observed market option price across all times $t$ and all options $\Omega_j$.
Let $f_{j, \lambda}(x,y,t)$ be the solution of the option pricing PDE \eqref{PDE} for option $\Omega_j$. The predicted option price $\hat{\Omega}_j(kT)$ for day $k$ becomes 
$f_{j, \lambda}(X_{kT}, Y_{kT}, kT)$. To compare this option price predictor to the observed option price $\Omega_j(kT)$, we compute the root mean squared prediction error $RMS_j(\lambda)$ defined by 
\begin{equation*}
RMS_j(\lambda)^2 = \frac{T}{\tau_j} \, \sum_{k=1}^{\tau_j/T} \left[ \, ( \, \hat{\Omega_j}(kT) - \Omega_j(kT) \, )^2 \, \right] 
\end{equation*}
Let $p_j$ be the median price of $\Omega_j$ over its lifetime. To combine option pricing accuracy across multiple options, introduce the relative size of pricing prediction errors $RMS_j(\lambda)/ p_j$ for each $j$. Consider any estimator $\hat{\lambda}$ of $\lambda$, computed from  the pool of options $\Omega_j$.  If $\hat{\lambda}$ takes the value $L$, we quantify the associated   \textit {prediction error} $\text{prederr}\,(L)$ by the \textit{median} of relative  pricing prediction errors over all benchmark options, so that 
\begin{equation*} 
\text{prederr} \, (L) = \; \text{median}\;_{j= 1 \ldots q} \left[ \; RMS_j(L)/ p_j \; \right]
\end{equation*}
We will hence compute  our estimator $\hat \lambda$  by minimizing the median pricing error $\text{prederr} \, (L) $ overall $L$, so that $\hat \lambda = L*$ where $L* > 0$  is determined  by 
\begin{equation*} 
 \text{prederr} \, (L*)= \text{arg} \, \min_{L > 0} \text{prederr} \,(L) 
\end{equation*}
Note that our approach to estimate $\lambda$ works just as well if we replace the median by the mean  of the $RMS_j / p_j$; but we prefer to use the median to gain in robustness to outliers.
\subsection{Numerical estimate of $\lambda$ for options written on S\&P 500}
\begin{figure}
\centering
\includegraphics[width=.9\textwidth, height=.4\textheight]{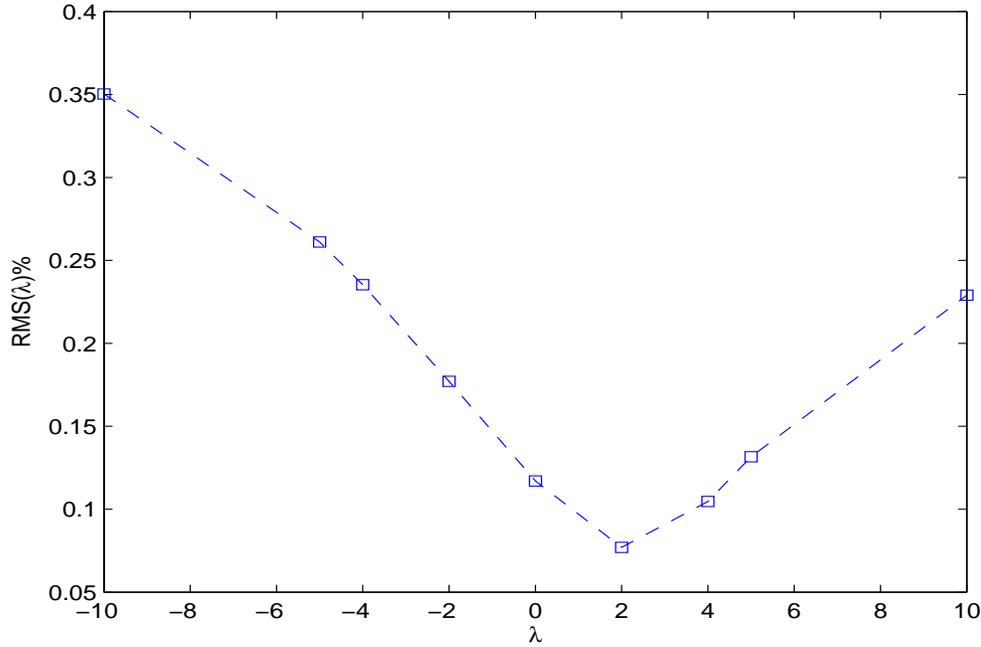}
\caption{For each potential value $\lambda$ of the market price of volatility risk, and each one of 16 benchmark option written on SPX, we compute the relative error of option pricing, and display the median $\text{prederr} \, (\lambda)$ of these 16 pricing errors. Note that $\text{prederr} \, (\lambda)$ reaches its minimum at $\lambda = 2.0$  which becomes our estimate  $\hat \lambda$ }. \label{lambda_new} 
\end{figure}

For the option pricing sensitivity results presented below, we have focused on the S\&P 500 index SPX and its approximate annualized volatility VIX in the first quarter of 2007. In the previous section, we have modeled the joint process $(SPX, VIX)$  by joint Heston SDEs with sub-sampling time $T=1/252$ and parameters estimated on the basis of 252 daily observations recorded in 2006.\\
To estimate $\lambda$ we have selected 16 benchmark options  $\Omega_j$ observed over 22 trading days. Eight of these options were maturing on Feb 17th 2007, with strike prices 1380, 1400, 1410, 1420, 1425, 1430, 1450, 1460. The other 8 options had the same strike prices but with maturity date Mar 17th 2007. \\
The risk free rate of return is set at $r=1\%$. Fix any tentative value $L$ of the unknown $\lambda$ As indicated above, for each option $\Omega_j$,  we solve the $\Omega_j$ pricing PDE to evaluate the mean squared difference $RMS_j(L)$ between real and computed $\Omega_j$ price over the 22 days option lifetime. We then compute the median 
$\text{prederr} \, (L) $ of the 16 relative pricing errors $RMS_j(L)/p_j$ where $p_j$ is the median price of $\Omega_j$ over its lifetime. 
The graph of $ \text{prederr} \,(L) $ is displayed in figure \ref{lambda_new} as a function of $L$, which reaches a minimum of $8\%$ for $L* = 2.0$. We thus estimate the unknown $\lambda$ by $\hat \lambda = L*= 2.0$ This estimate of $\lambda$ then enables us to compute all the  of option pricing sensitivities we present for options based on S\&P 500.

\subsection{Numerical computing of option price sensitivities} \label{benchmark_option}
%%%%%%%%%%%%%%
We study two benchmark European options $\Omega_1$ and $\Omega_2$ written on the SPX index, maturing at $ 63$ days and $126$ days, and with identical strike price $K = 1380$. These options were actively traded during the 1st quarter of 2007. \\
The time between daily observations for the underlying asset model has been conventionally set at $T=1/252$, so in option pricing PDEs \eqref{PDE}, the maturity $\tau$ must be $\tau_1 =63/252 =0.25$ for $\Omega_1$ and $\tau_2 = 126/252 = 0.5$ for $\Omega_2$.\\
The risk free rate of return is set at $r= 1\%$. \\
As detailed in section \ref{lambdaest} above, we  assume that the market price $\lambda$ of volatility risk is constant, and we estimate it by analysis of 16 other options written on SPX, which yields the  estimate $\hat \lambda = 2$.\\
In the first quarter of 2007, $SPX$ ranged from 1370 to 1460 with median 1426, and $VIX$ ranged from $10\%$ to $20\%$ with median $11\%$. We use these values to determine realistic domains in $R^3 $ for the triple $X_t,Y_t, t$ defined by $0 < x < 2800 $, $0 < y < 1$, $ 0 < t< \tau$, with $\tau = \tau_1 = 0.25$ for $\Omega_1$ and $\tau = \tau_2 = 0.5$ for $\Omega_2$. \\
In these two domains, we need to solve the option pricing PDE \eqref{PDE} for $g(x,y,t) = f(x,y,\tau -t)$ with boundary conditions \eqref{f.boundaryinit}-\eqref{f.boundaryneumy}.\\
After checking numerically that for $100< x <2800$ option prices are not significantly affected when the initial boundary $x=0$ is shifted to the position $x= 100$, we did implement this shift for substantial CPU reduction. \\
%The change of variable $x \rightarrow \log(x)$ in PDE \eqref{PDE} and its boundary conditions transform the computing domains into
%$$
%7 < \log(x) < 8 , \quad 0 <y <1 , \quad 0 <t < \tau , \;\; \text{with} \; \tau_1 = 0.25 \;\; \text{and} \; \tau_2 = 0.50
%$$
The computing domain is then
$$
100 < x <2800 , \quad 0 <y <1 , \quad 0 <t < \tau , \;\; \text{with} \; \tau_1 = 0.25 \;\; \text{and} \; \tau_2 = 0.50.
$$
We discretize this domain by a grid $G_1$ of size $90 \times 80 \times 63$ for $\Omega_1$, and a grid $G_2 $ of size $90 \times 80 \times 126$ for $\Omega_2 $. As indicated in section \ref{numerical_imp}, we then solve the four PDEs \eqref{sens_kappa} -- \eqref{sens_rho}, verified by option price partial derivatives with respect to $\kappa, \theta, \gamma, \rho $, to obtain the vector $SEN(\Theta)$ of option pricing sensitivities $\; ( Sen_{\kappa} , Sen_{\theta} , Sen_{\gamma} , Sen_{\rho} )$. \\
By refining the grids $G_1$ and $G_2$, we have numerically validated that these discretizations were dense enough to accurately solve all the necessary PDEs.\\
Since the unknown $\Theta$ may essentially be any point of the high probability localization box $J \subset R^4$ defined above in \ref{benchmark_option}, the individual impacts on option pricing of parameter estimation errors, denoted $ \varepsilon_{\kappa} , \; \varepsilon_{\theta} , \; \varepsilon_{\gamma} , \; \varepsilon_{\rho} $ in equation \eqref{singleimpact}, are then evaluated by their upper bounds for $\Theta$ in a large finite subgrid of $J$.\\
On a standard desktop PC, solving all the option pricing and sensitivity PDEs as above required for each $\Theta$ a CPU-time of 1 minute for option $ \Omega_1$ and 2 minutes for option $\Omega_2$. When the grid size $m \times n \times p$ increases, these CPU times are roughly linear in the grid time size $p$ and behave like low degree polynomials in the grid spatial size $m n$; indeed, a key computational cost is the LU decomposition and backward substitution for a single $m n\times m n$ matrix. \\
\section{Impact of parameter estimation errors on option pricing} 
\begin{figure}[h]
\subfigure
{\includegraphics[width=.8\textwidth]{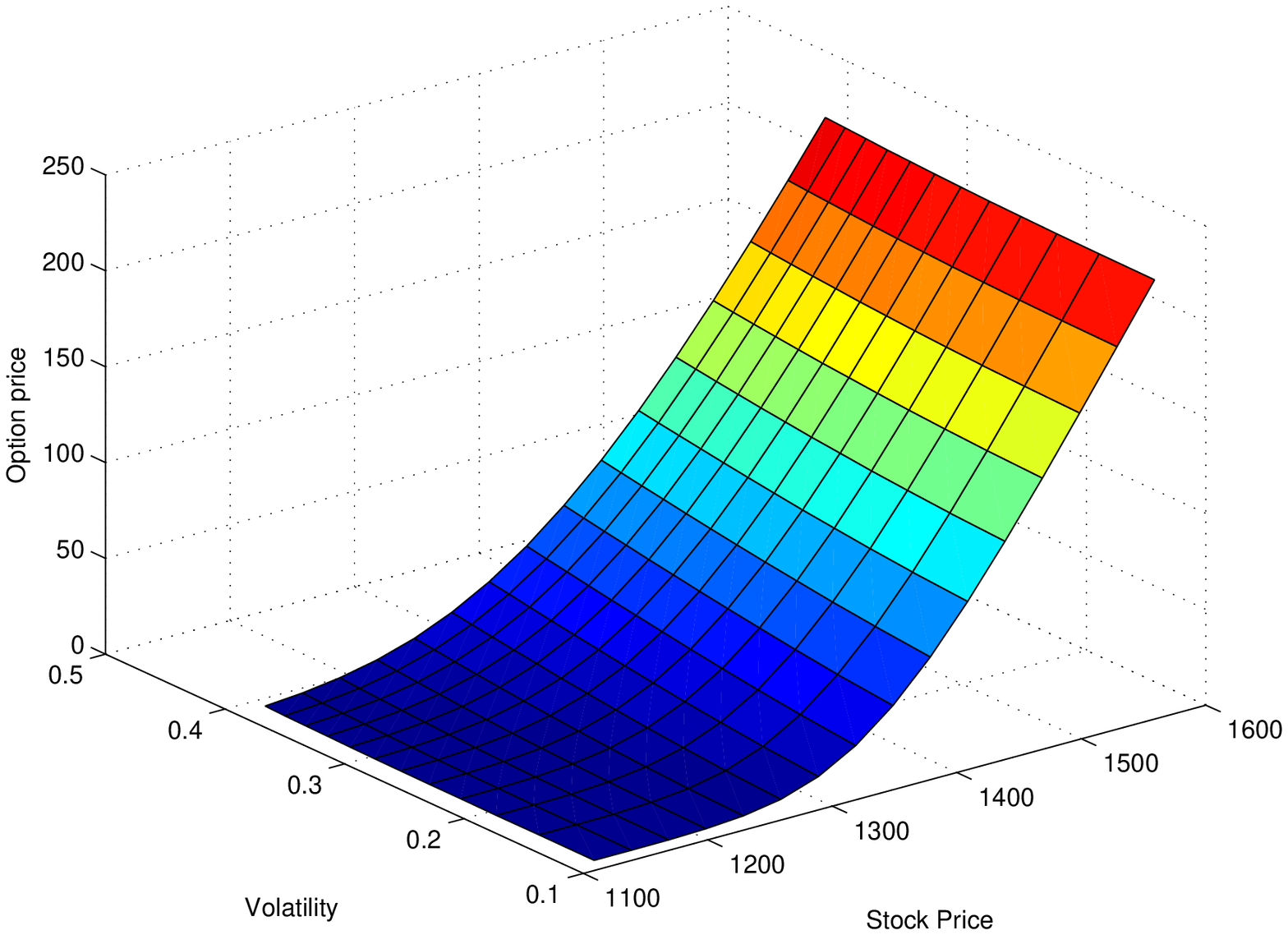}} 
\subfigure
{\includegraphics[width=.8\textwidth]{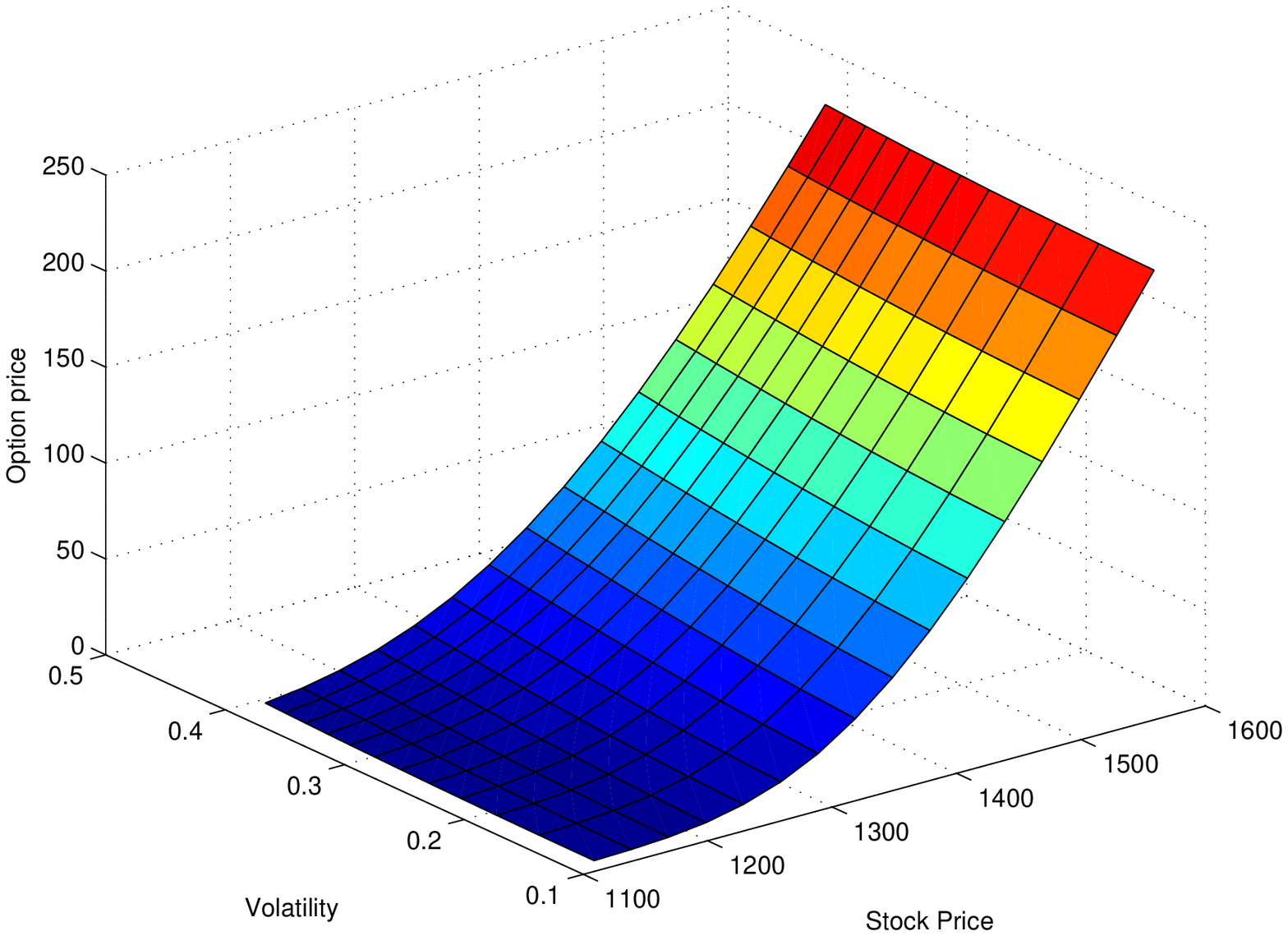}} 
\caption{Computed price of options $\Omega_1$(top) and $\Omega_2$(bottom) written on $SPX$. Option prices are computed at option creation time; the horizontal coordinates $x$ and $\sqrt{y}$ denote the SPX price and its approximate volatility VIX.} \label{priceset1}
\end{figure}

\begin{figure}
\subfigure[$\Omega_1$]
{\includegraphics[width=.45\textwidth]{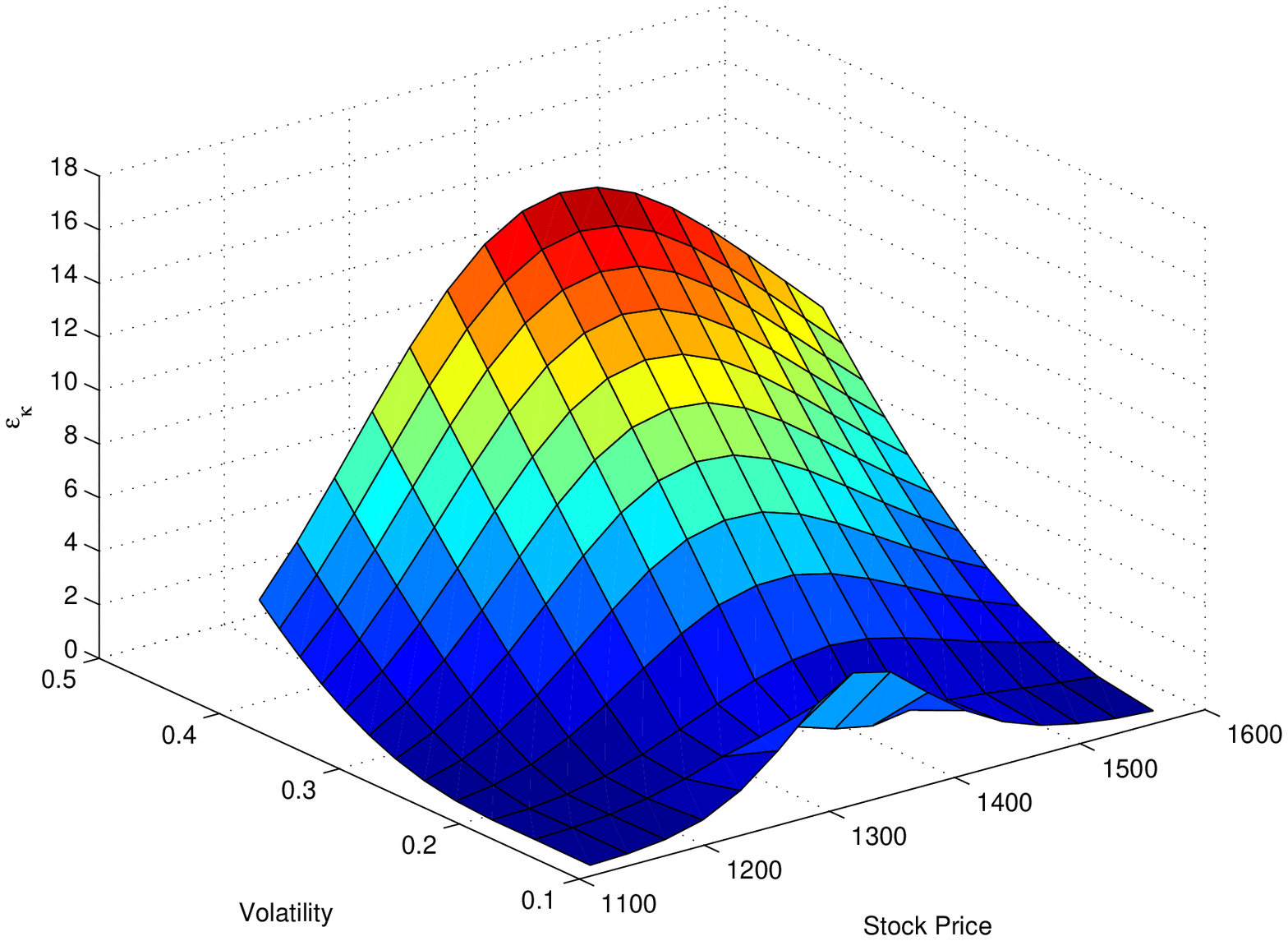}} 
\subfigure[$\Omega_2$]
{\includegraphics[width=.45\textwidth]{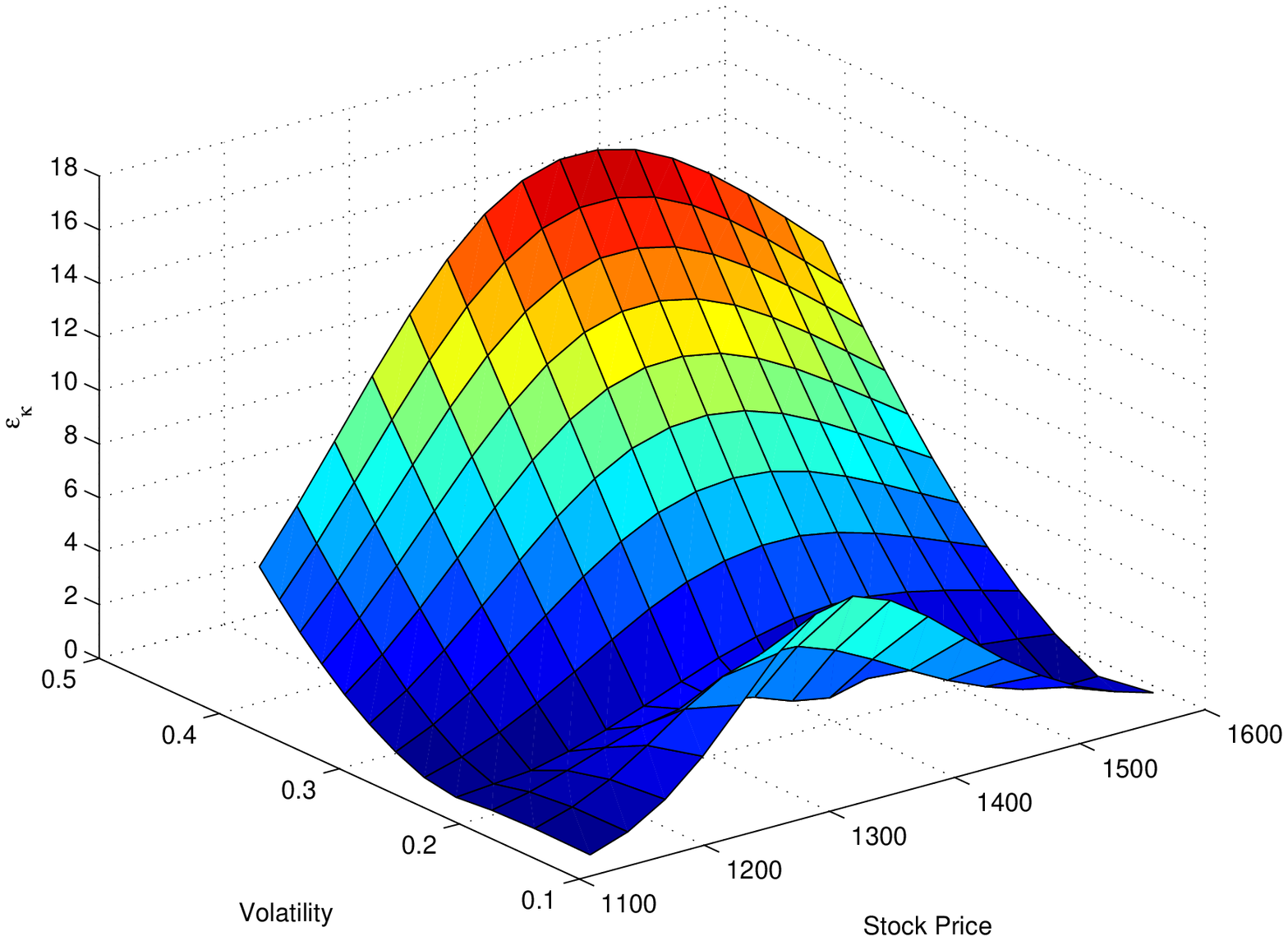}}
\caption{Option pricing errors due to estimation errors on $\kappa$, computed at option creation time.}\label{impactkappa}
\end{figure}

\begin{figure}
\subfigure[$\Omega_1$]
{\includegraphics[width=.45\textwidth]{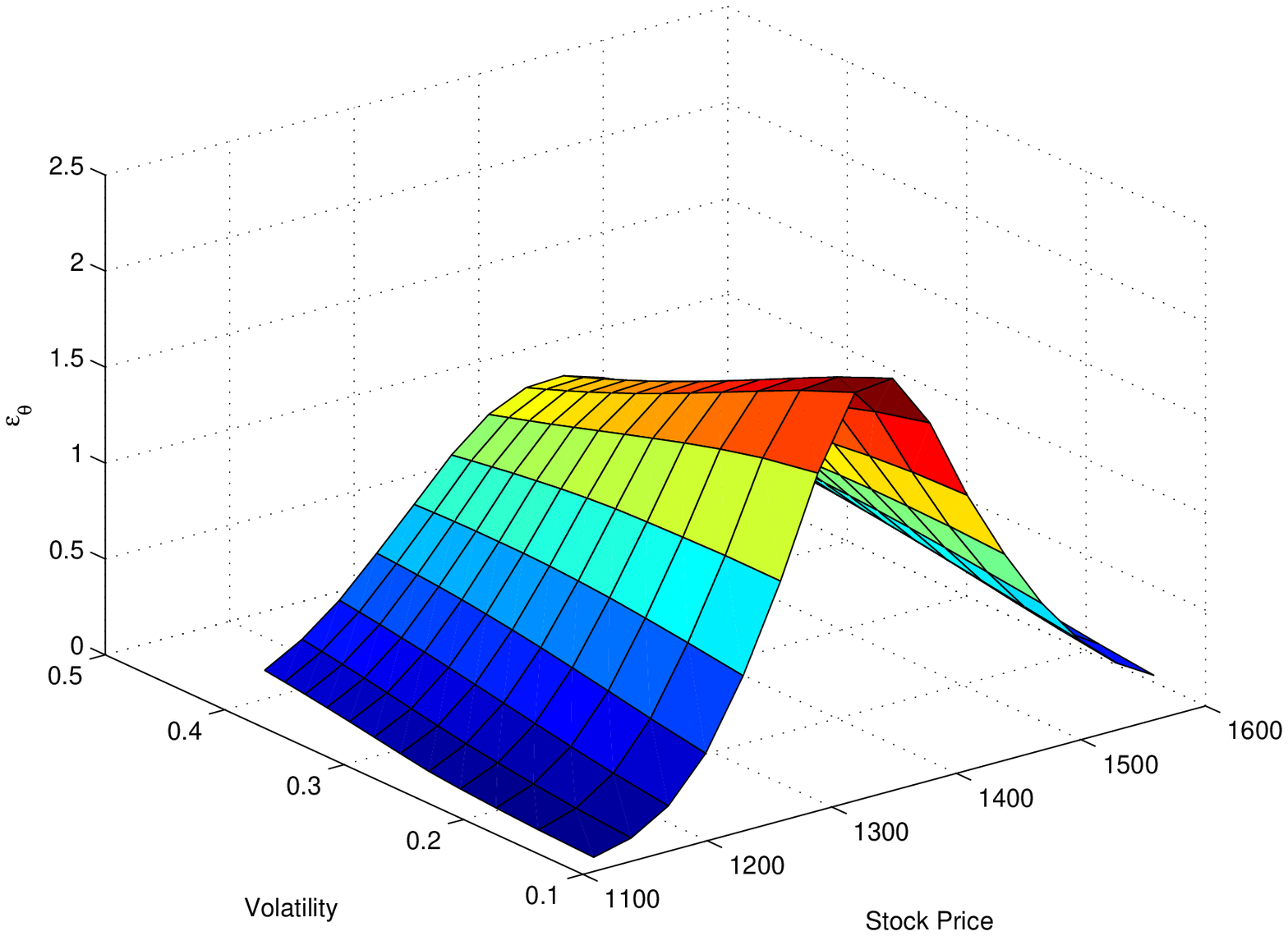}} 
\subfigure[$\Omega_2$]
{\includegraphics[width=.45\textwidth]{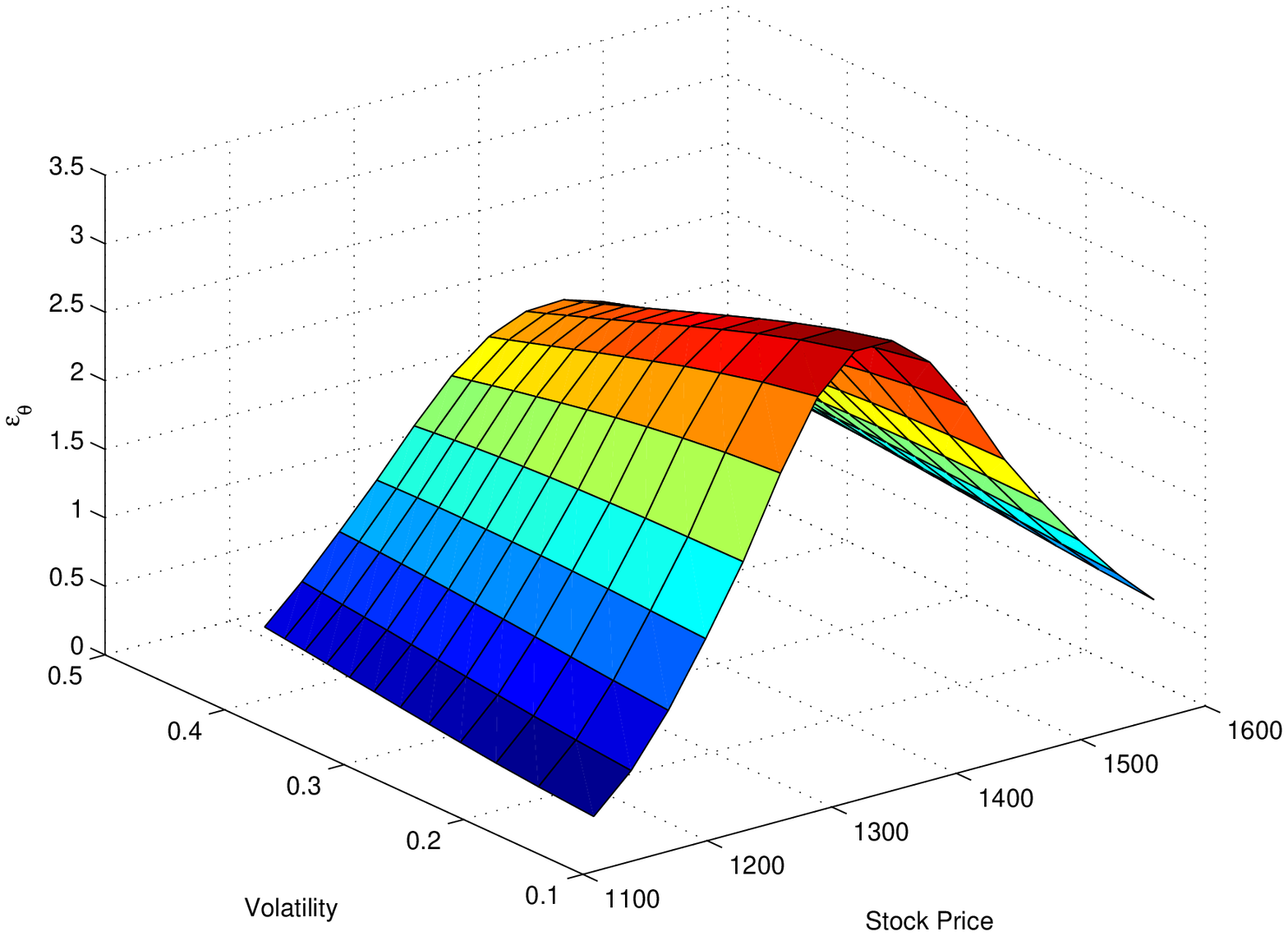}}
\caption{Option pricing errors due to estimation errors on $\theta$, computed at option creation time.}\label{impacttheta}
\end{figure}

\begin{figure}
\subfigure[$\Omega_1$]
{\includegraphics[width=.45\textwidth]{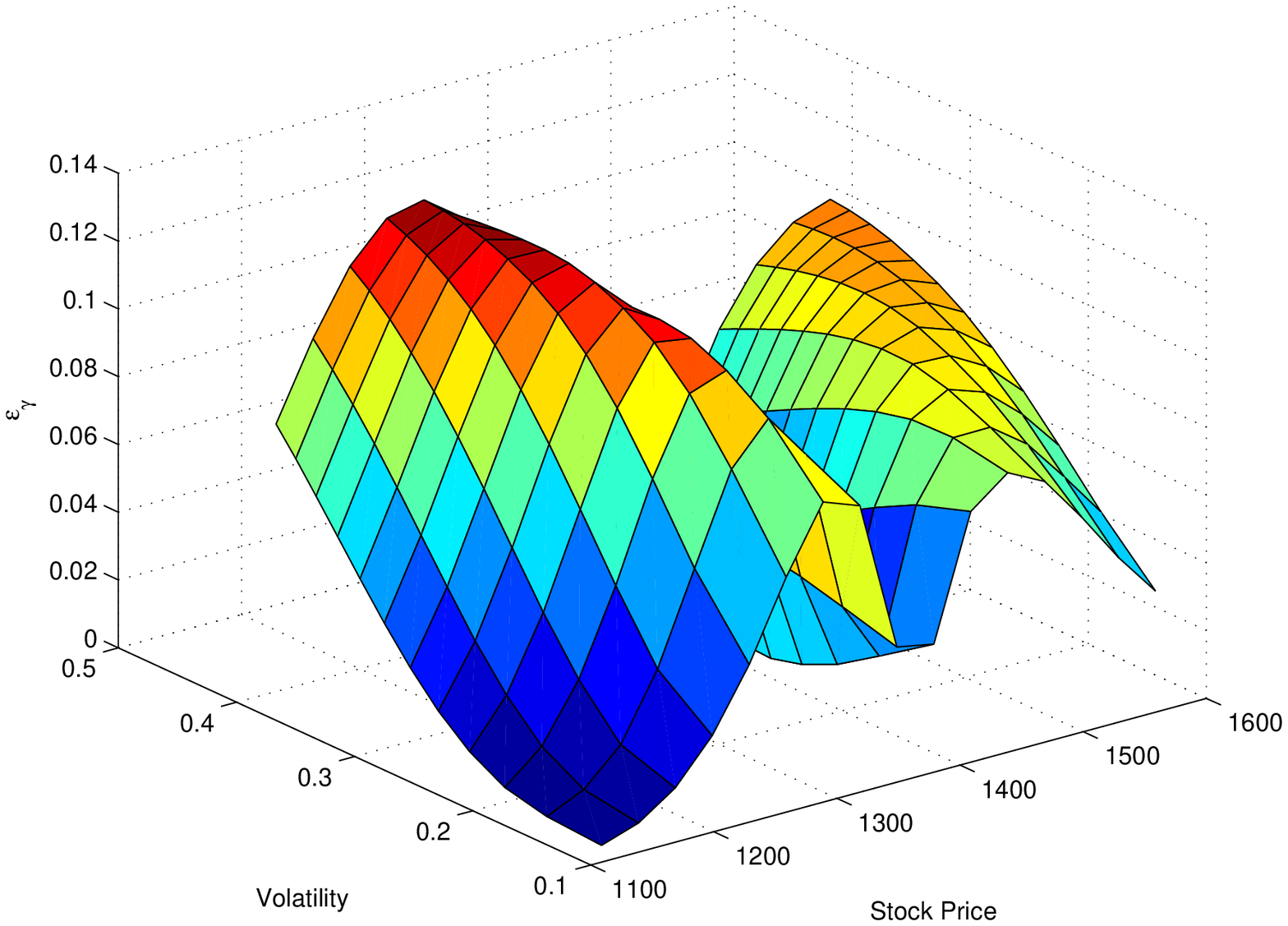}} 
\subfigure[$\Omega_2$]
{\includegraphics[width=.45\textwidth]{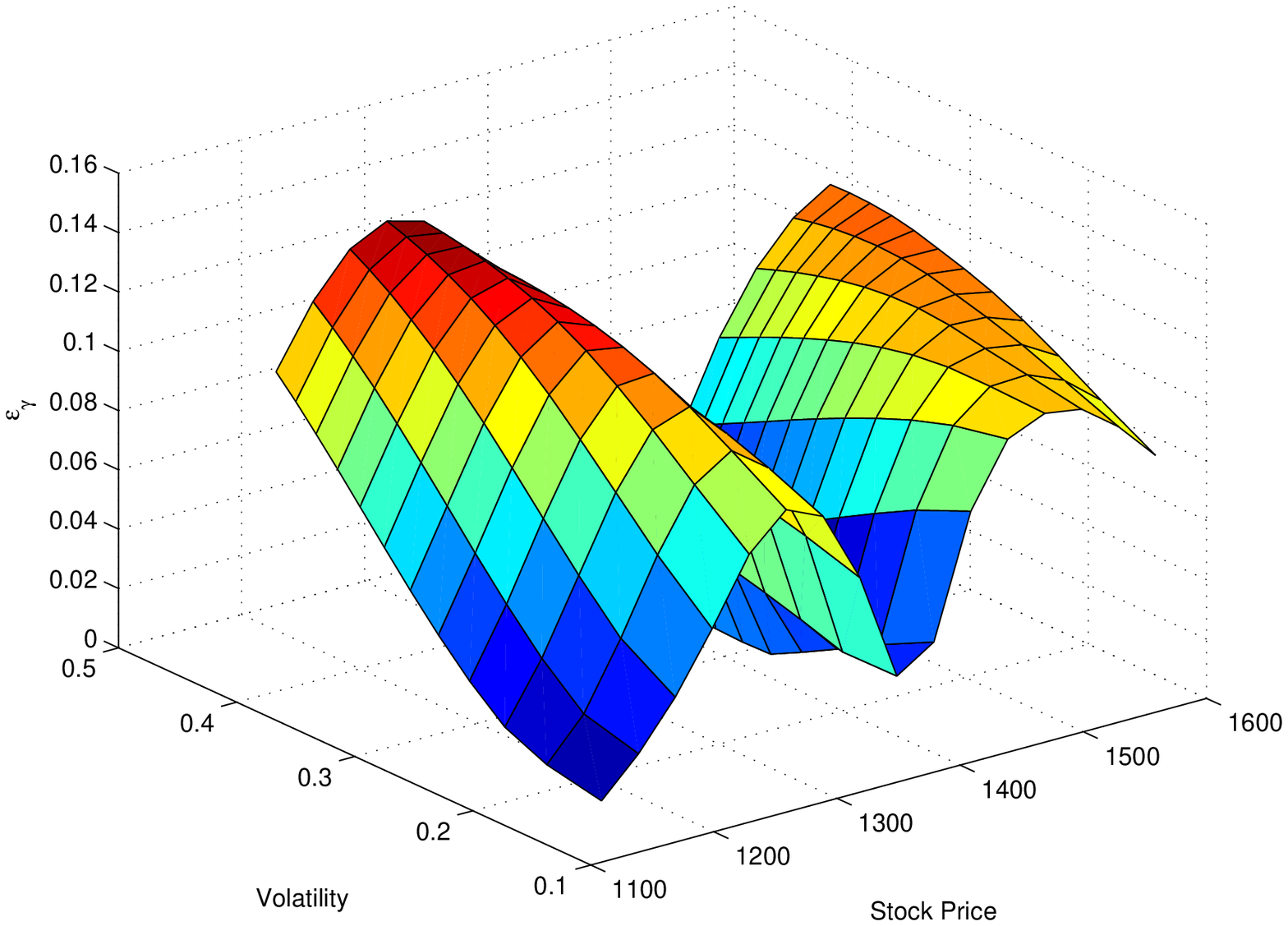}}
\caption{Option pricing errors due to estimation errors on $\gamma$, computed at option creation time.}\label{impactgamma}
\end{figure}

\begin{figure}
\subfigure[$\Omega_1$]
{\includegraphics[width=.45\textwidth]{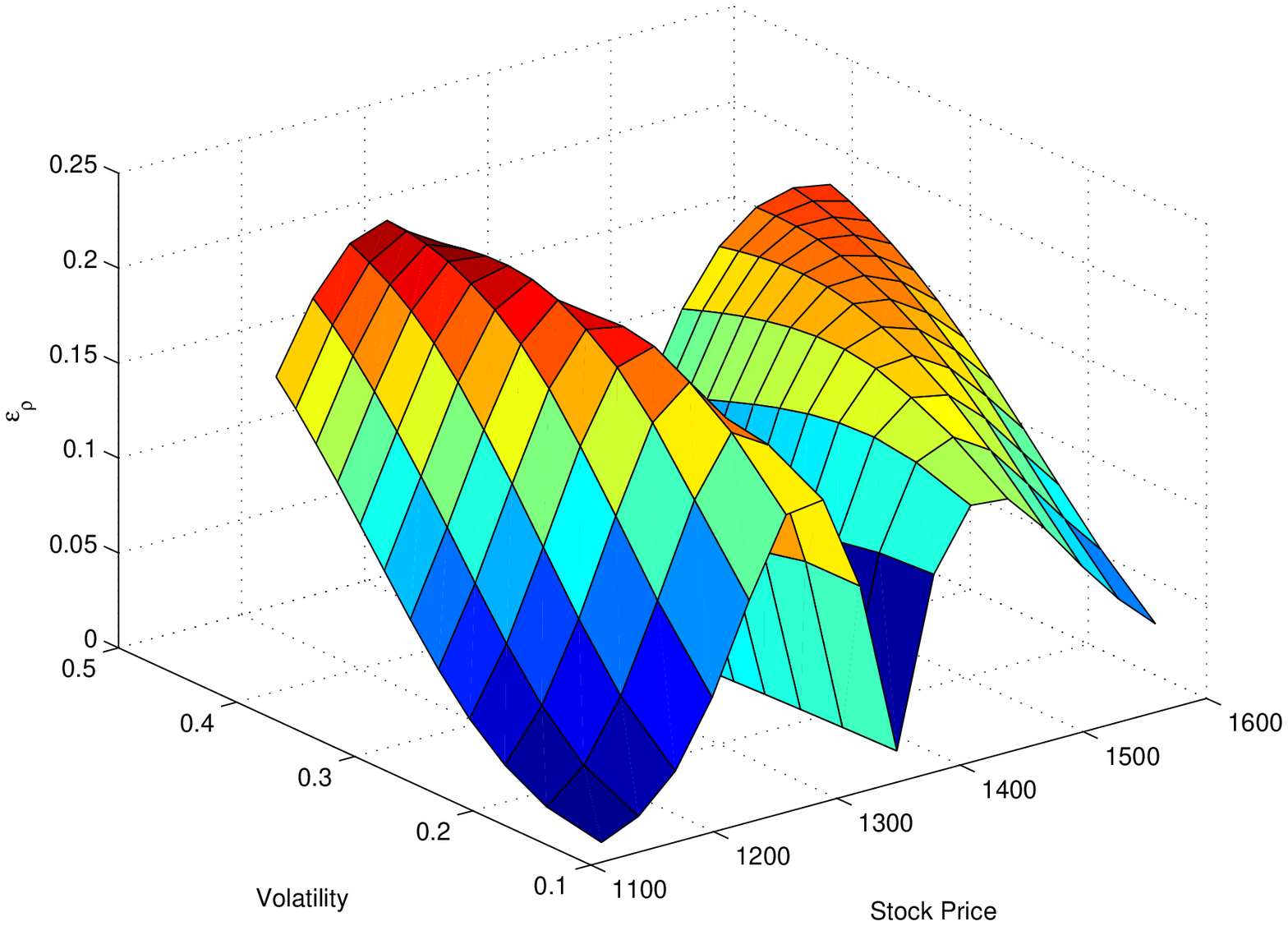}} 
\subfigure[$\Omega_2$]
{\includegraphics[width=.45\textwidth]{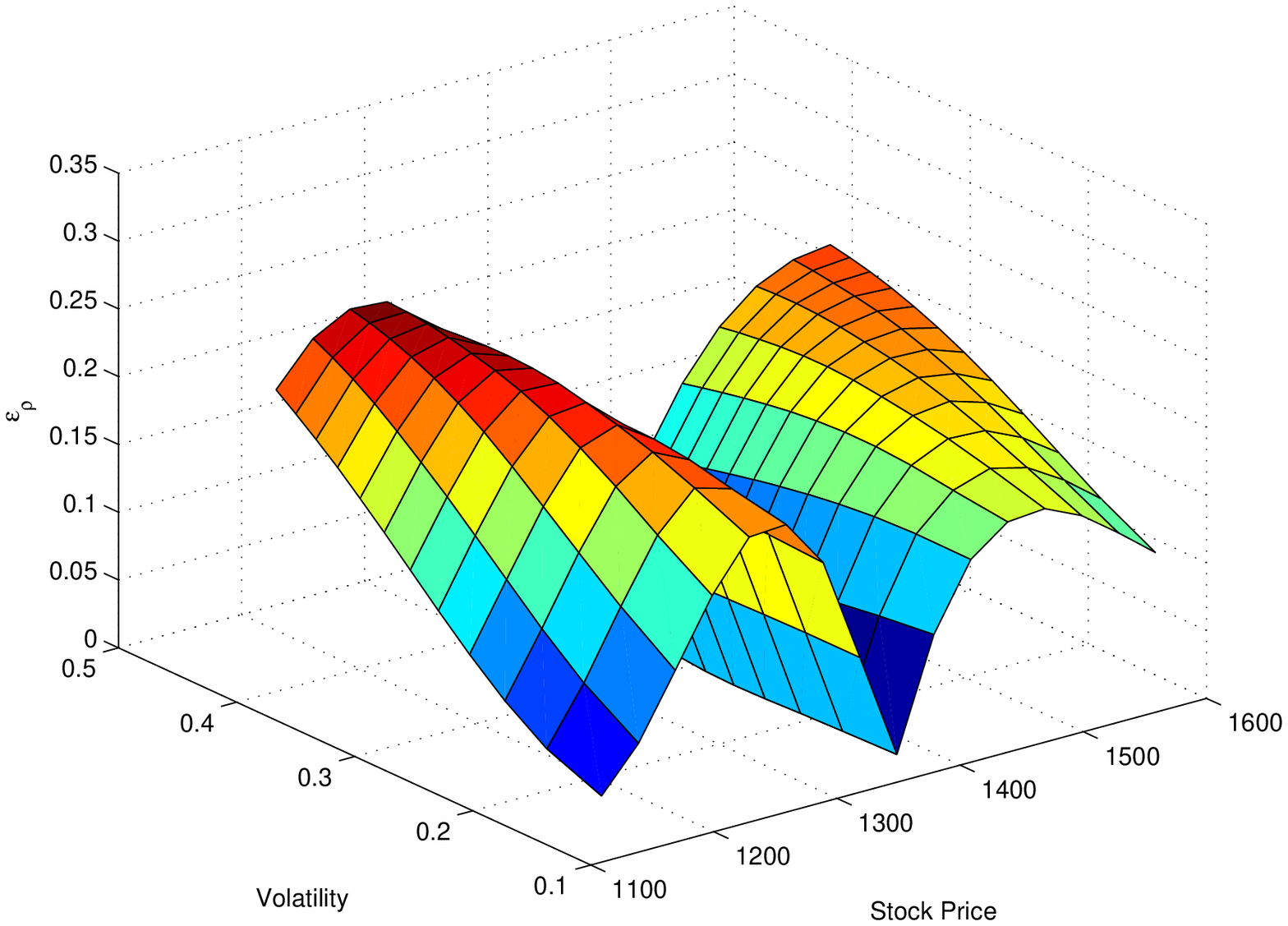}}
\caption{Option pricing errors due to estimation errors on $\rho$, computed at option creation time.} \label{impactrho}
\end{figure}

\begin{figure}
\includegraphics[width=.8\textwidth]{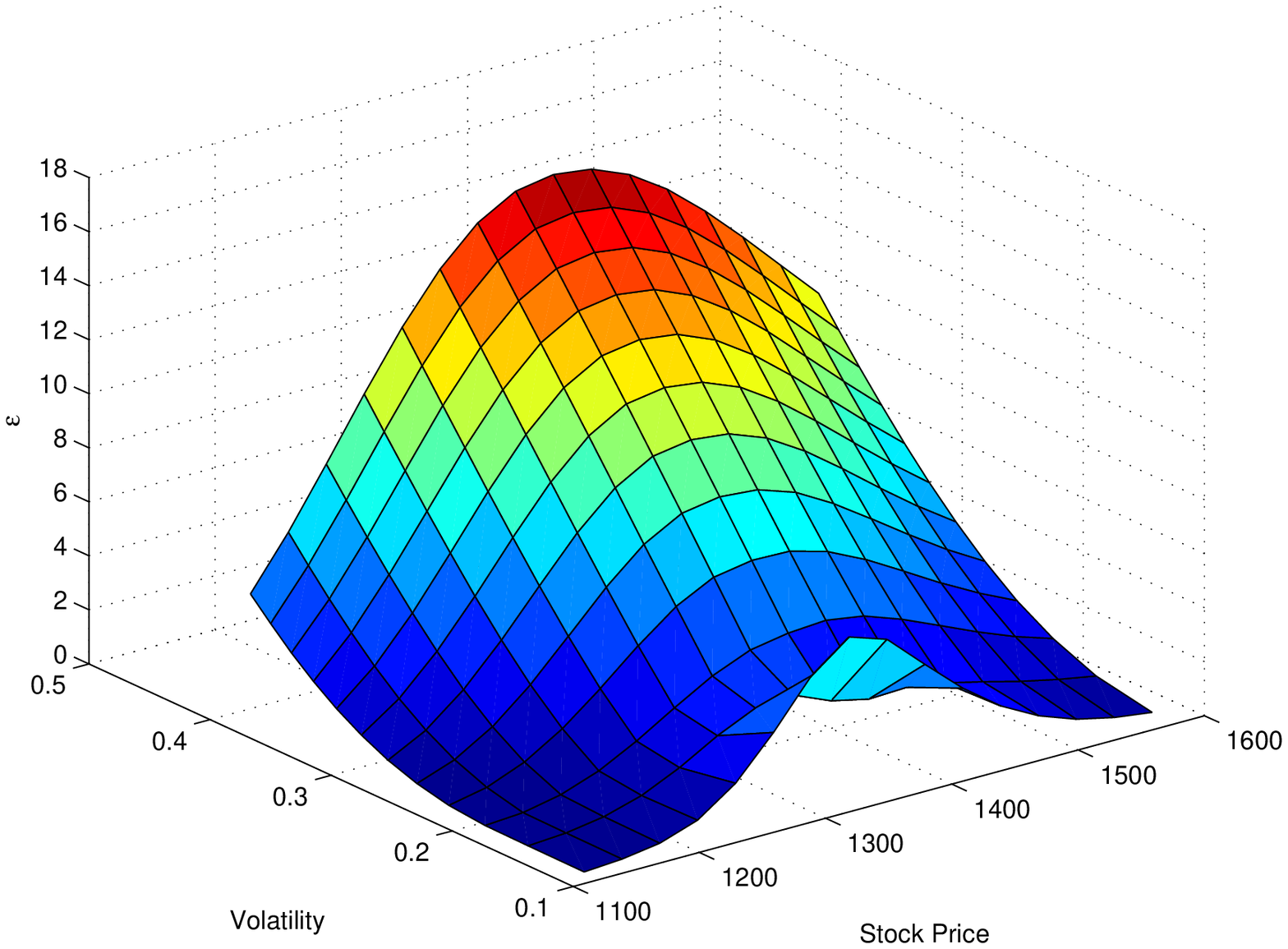}
\caption{Global option pricing errors in dollars for option $\Omega_1$, computed at option creation time.}\label{global_error_O1}
\end{figure}

\begin{figure}
\includegraphics[width=.8\textwidth]{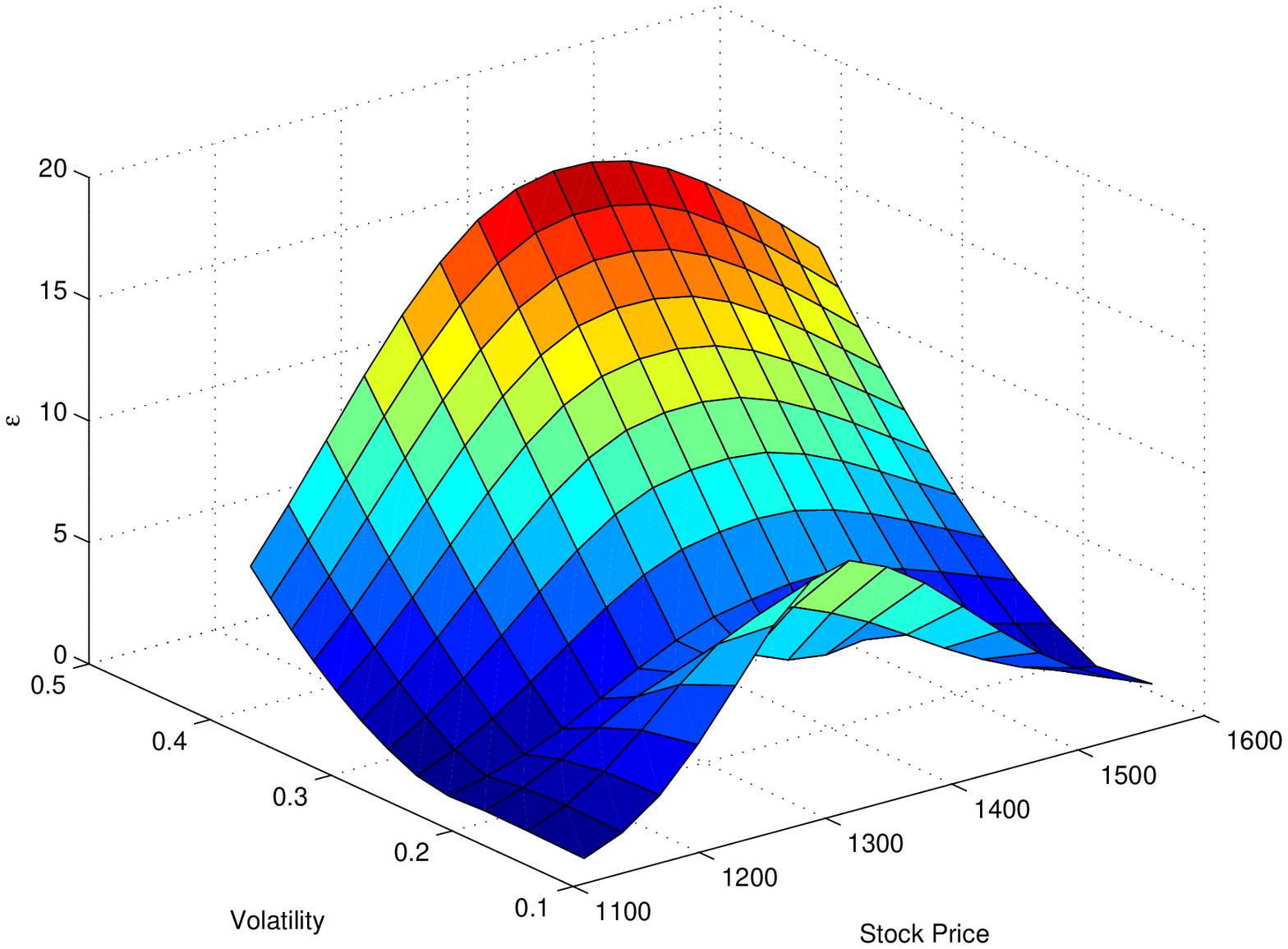}
\caption{Global option pricing errors in dollars for option $\Omega_2$, computed at option creation time.}\label{global_error_O2}
\end{figure}

\begin{figure}
\includegraphics[width=.8\textwidth]{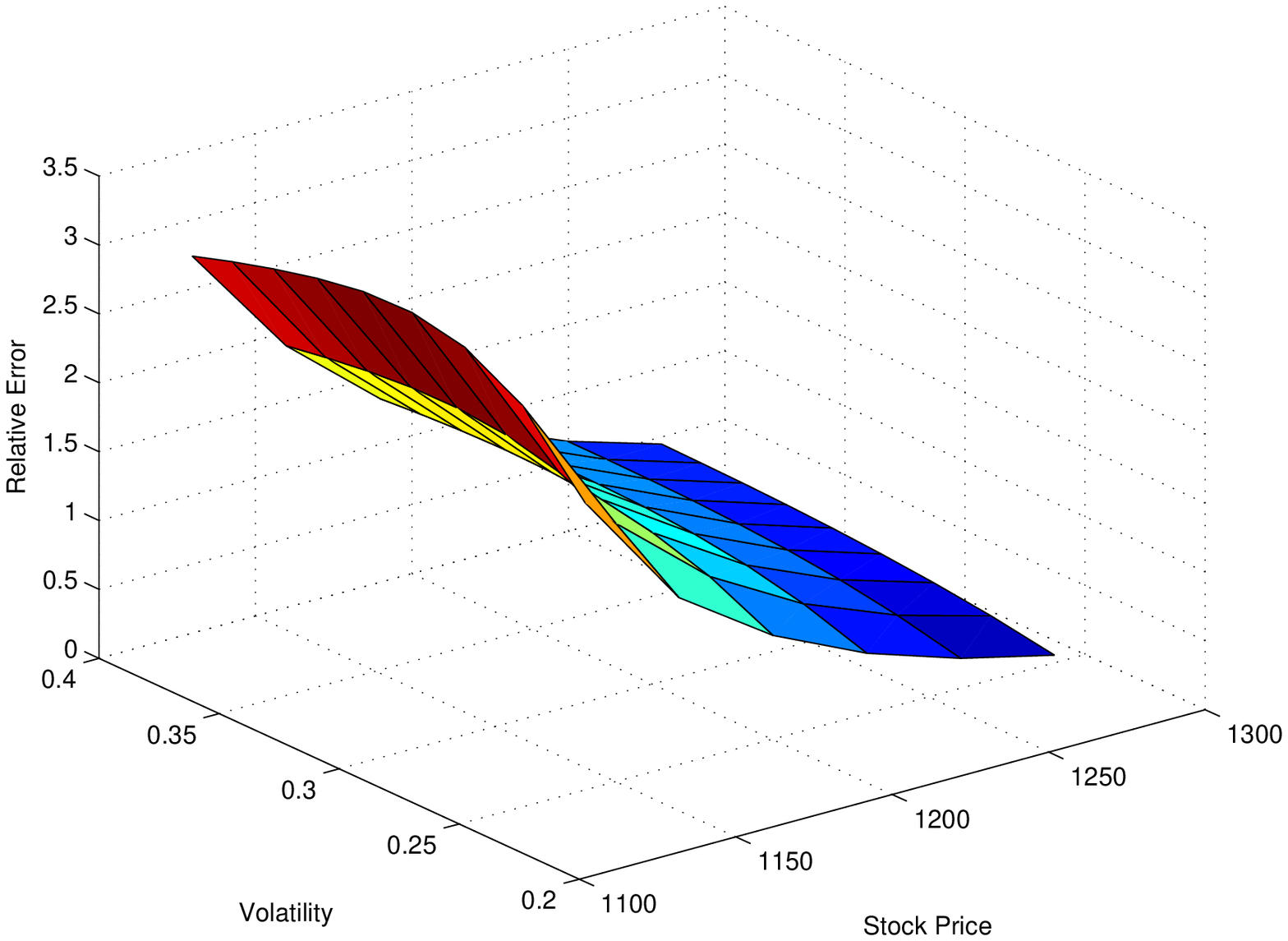}
\caption{Relative Errors on option prices for option $\Omega_1$, computed at option creation time.}\label{relative_error_O1}
\end{figure}

\begin{figure}
\includegraphics[width=.8\textwidth]{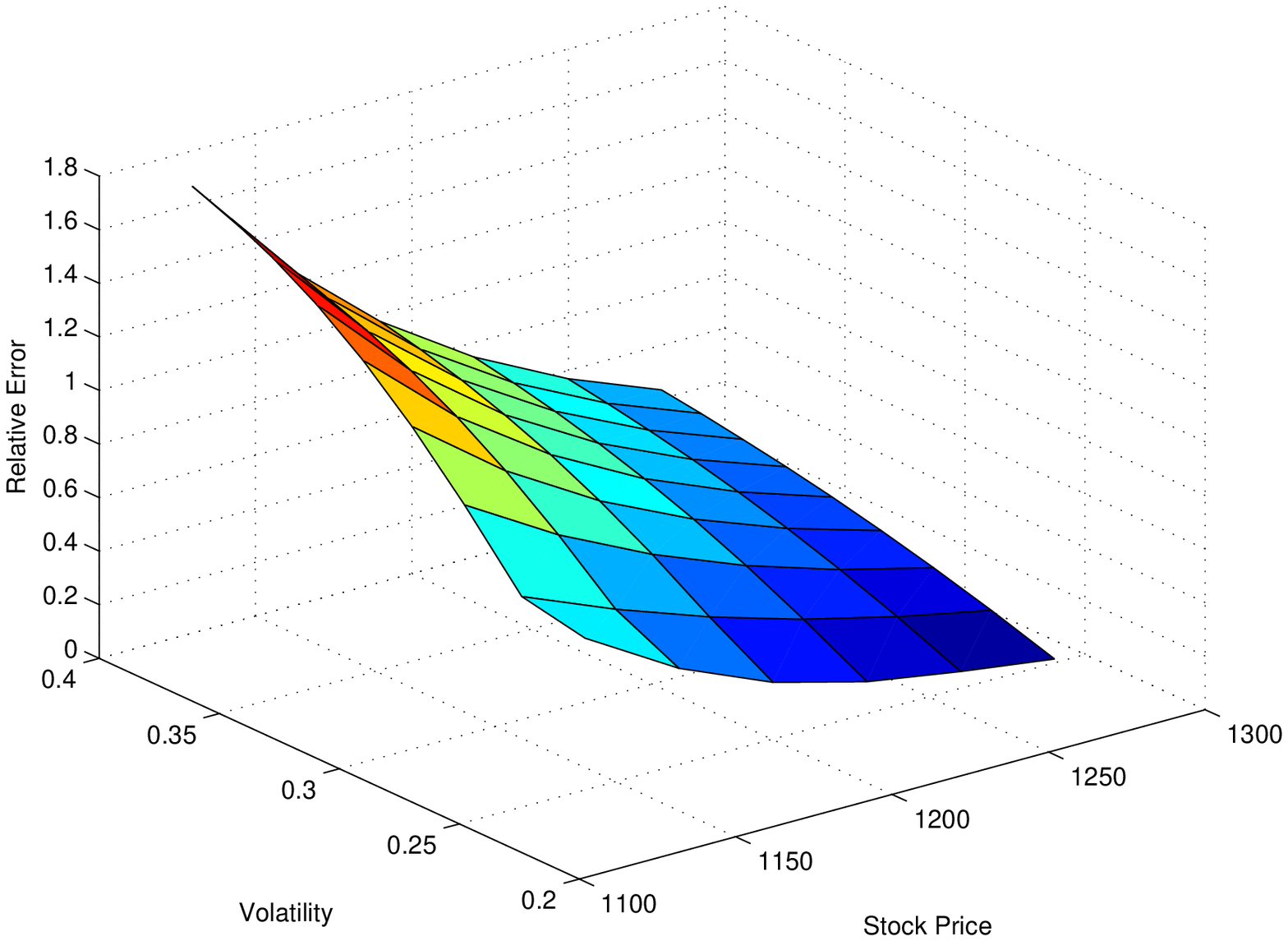}
\caption{ Relative Errors on option prices for option $\Omega_2$, computed at option creation time.}\label{relative_error_O2}
\end{figure}

We now present numerical results for the two European options $\Omega_1,\Omega_2$ written on the SPX index, with strike price and maturities that were actually traded in 2007. The 3D-graphs in Fig. \ref{priceset1} display the prices of options $\Omega_1,\Omega_2$ computed at option creation time, as functions of current SPX price $x \in [1120,1570]$ and VIX value with $ \sqrt{y} \in [11\% , 38\%]$. 
For options $\Omega_1$ and $\Omega_2$, the figures Fig. \ref{impactkappa}, \ref{impacttheta}, \ref{impactgamma}, \ref{impactrho} display the individual impacts $ \varepsilon_{\kappa} , \; \varepsilon_{\theta} , \; \varepsilon_{\gamma} , \; \varepsilon_{\rho} $ of parameter estimation errors on option pricing. These separate error impacts on option pricing are computed by equation \eqref{singleimpact}, and are displayed in our 3D-graphs as functions of the SPX price $1120 < x < 1570 $ and of its volatility, covering deep-out-of-the money cases as well as deep-in-the money cases. \\ 
Note that the estimation errors on $\kappa$ and $ \theta $ induce option pricing errors which tend to be decreasing functions of $|x-K|$ where $x= SPX_t$ and $K$ is the option strike price. Since the individual impacts of estimation errors on option price are clearly stronger for $\kappa$ and $ \theta $ than for $\gamma$ and $ \rho$, we see that option pricing tends to be more sensitive to parameter estimation errors for options close to the money than for options far from the money.\\
The bound on the global option pricing error $\varepsilon$ induced by the combined effects of all parametric estimation errors was computed as the upper bound of the right-hand-side of equation \eqref{eps2} for $\Theta$ in a large finite subgrid of the localization box $J$ and displayed in Fig. \ref{global_error_O1} and \ref{global_error_O2} as a function of current asset price and volatility. \\
Relative global option pricing errors are defined by $\varepsilon (x,y,t) / f(x,y,t)$ and are displayed in Fig. \ref{relative_error_O1} and \ref{relative_error_O2} for $1120 < x < 1270,\;\; 22\% < y < 38\%$.

%%%
%%%
%%%

%%%
\subsection{Impact on option price of estimation errors in $\lambda$ } 
\begin{figure}
\subfigure[$\Omega_1$]
{\includegraphics[width=.45\textwidth]{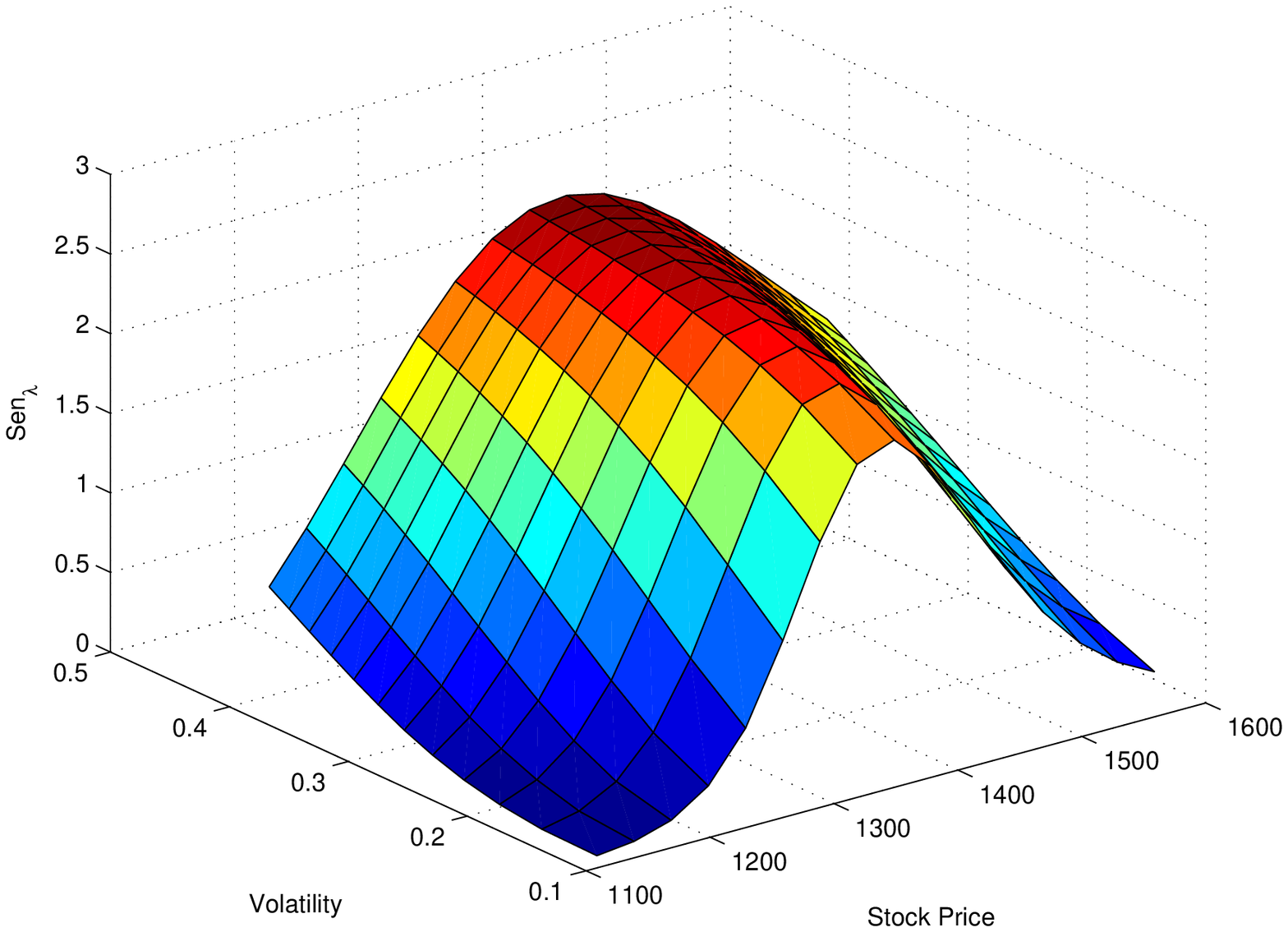}} 
\subfigure[$\Omega_2$]
{\includegraphics[width=.45\textwidth]{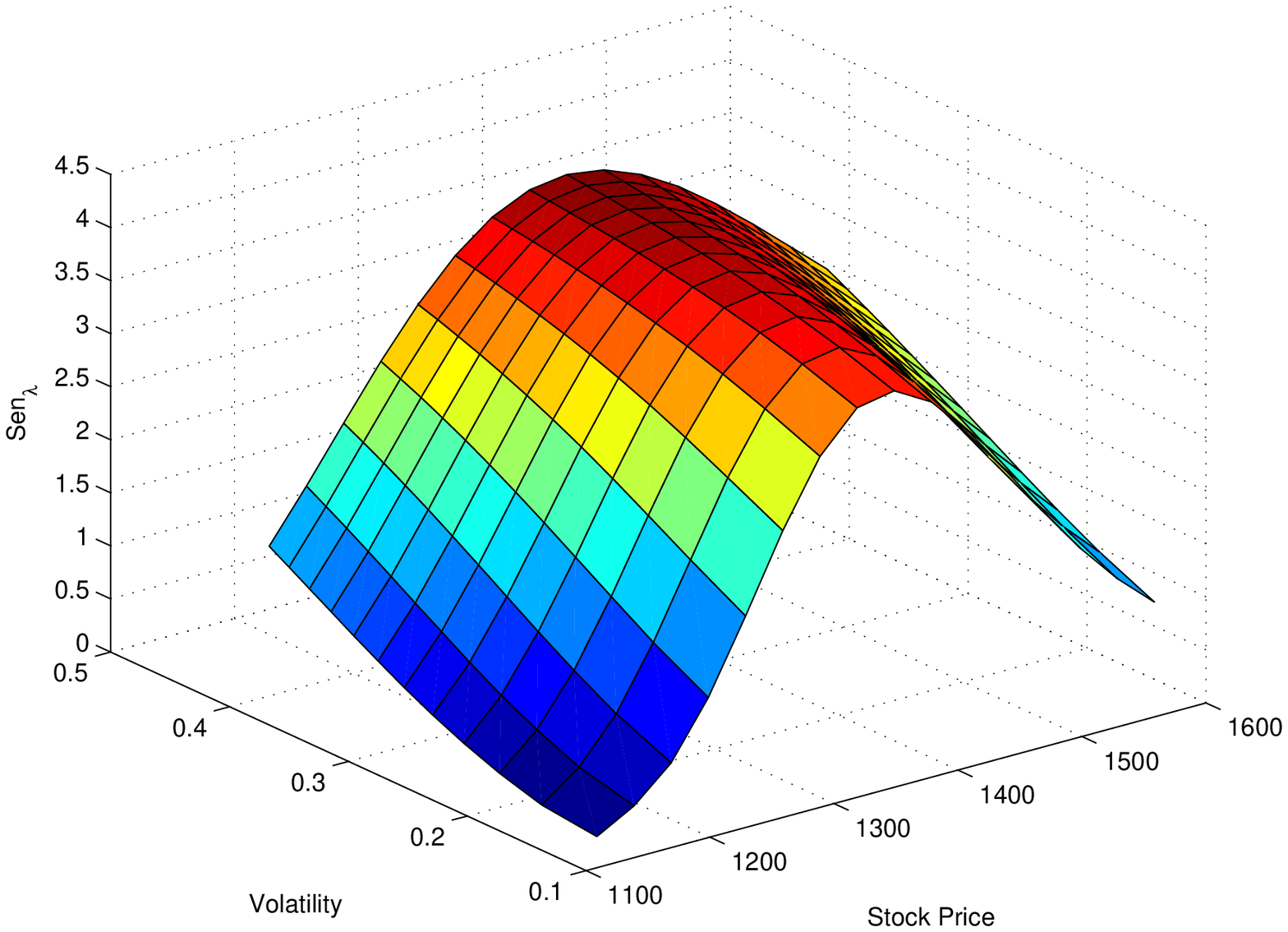}}
\caption{Option price sensitivity with respect to $\lambda$, computed at option creation time.}\label{sens_lam}
\end{figure}

Our estimate $\hat\lambda = 2.0$ was derived by minimizing the median option pricing error, where the median was computed over a set $K$ of 16 benchmark options $\Omega_j$ written on SPX within a short trading period. To evaluate the error of estimation $s_{\lambda}$ affecting $\hat\lambda$, we implement a rough ``bootstrap" evaluation as follows. For each subset $Q $ of 12 options arbitrarily selected in $K$, one can as above compute $prederr_Q(\lambda)$ as the median pricing accuracy over the 12 options in $Q$ and then minimize $prederr_Q(\lambda)$ in $\lambda$, which yields another estimate $\hat\lambda(Q)$ of $\lambda$.\\
The average of the $\frac{16!}{4! 12!}$ shifts $\; | \hat\lambda(Q) - \hat\lambda | \;$ is a rough evaluation of the error $s_{\lambda}$ affecting the estimate $ \hat\lambda $. This procedure provides here the value $s_{\lambda} \sim 0.5 $.\\
The sensitivity $Sen_{\lambda} = | \partial_{\lambda}f(x,y,t) |$ of the option price $f(x,y,t)$ to errors affecting $\lambda$ is computed by solving the adequate PDE as indicated in section \ref{sensitivity}. Fig. \ref{sens_lam} displays the sensitivities of options $\Omega_1$ and $\Omega_2$ with respect to $\lambda$, computed at option creation time.
\section{Conclusion}\label{concl}
Our main goal was to quantify the impact on European call option pricing of the parameter estimation errors generated by fitting Heston joint SDEs to market data. Since volatility data are unobservable we consider that they are systematically estimated  either by realized volatilities or by implied volatilities. \\ 
%On the theoretical level we tend to favor adequately parameterized realized volatility estimates, since in several companion studies,  we have  %characterized the stability of large classes of Heston  parameter estimators when one replaces true volatilities by realized volatilities. 
We have developed and numerically tested an impact quantification technique combining the computation of consistent estimators for the underlying Heston SDEs parameters, the evaluation of root mean squared accuracy for these estimators, and the numerical solution of six parabolic partial differential equations in $\mathbb{R}^2$, namely the option pricing PDE and the PDEs verified by the partial derivatives of option price with respect to model parameters. \\
We have also derived and implemented an algorithm to estimate the market price of volatility risk by analysis of multiple benchmark options  \\
We have tested our approach by numerical fitting of Heston joint SDEs to the 252 daily data recorded in 2006 for the S\&P 500 and VIX indices, and by studying several European call options written on S\&P 500. For two such options, we compute and display the average option pricing shifts induced by errors of estimation on the SDEs coefficients, as well as by approximation errors for the market price of volatility risk. Since our algorithmic implementations are fairly fast on a standard laptop, our study strongly suggests that quantification of errors induced on option pricing by model parameters estimation errors should not be neglected, and could be systematically computed when option pricing is performed after fitting joint Heston SDEs to daily or intraday market data. Our numerical results indeed show that Heston SDEs fitting to a year of daily data can generate sizeable inaccuracies for European option pricing.\\
We expect that our methods will perform just as well for American options as for European options.  We also plan to extend our approach to develop fast accuracy monitoring algorithms for portfolio hedging.
\bibliographystyle{rQUF}
\bibliography{refs}
\end{document}